\documentclass{article}

\usepackage{arxiv}

\usepackage[utf8]{inputenc} 
\usepackage[T1]{fontenc}    
\usepackage{lmodern}        
\usepackage{hyperref}       
\usepackage{url}            
\usepackage{booktabs}       
\usepackage{amsfonts}       
\usepackage{nicefrac}       
\usepackage{microtype}      
\usepackage{graphicx}

\title{Principal Component Analysis and biplots. A Back-to-Basics
Comparison of Implementations}

\author{
    Ettore Settanni
   \\
    Institute for Manufacturing \\
    University of Cambridge \\
  Cambridge, UK \\
  \texttt{\href{mailto:E.Settanni@eng.cam.ac.uk}{\nolinkurl{E.Settanni@eng.cam.ac.uk}}} \\
  }

\usepackage{color}
\usepackage{fancyvrb}

\DefineVerbatimEnvironment{Highlighting}{Verbatim}{commandchars=\\\{\}}
\usepackage{framed}
\definecolor{shadecolor}{RGB}{248,248,248}
\newenvironment{Shaded}{\begin{snugshade}}{\end{snugshade}}

\newcommand{\AttributeTok}[1]{\textcolor[rgb]{0.13,0.29,0.53}{#1}}

\newcommand{\ConstantTok}[1]{\textcolor[rgb]{0.56,0.35,0.01}{#1}}
\newcommand{\ControlFlowTok}[1]{\textcolor[rgb]{0.13,0.29,0.53}{\textbf{#1}}}

\newcommand{\DecValTok}[1]{\textcolor[rgb]{0.00,0.00,0.81}{#1}}

\newcommand{\FunctionTok}[1]{\textcolor[rgb]{0.13,0.29,0.53}{\textbf{#1}}}

\newcommand{\NormalTok}[1]{#1}

\newcommand{\OtherTok}[1]{\textcolor[rgb]{0.56,0.35,0.01}{#1}}

\newcommand{\SpecialCharTok}[1]{\textcolor[rgb]{0.81,0.36,0.00}{\textbf{#1}}}

\newcommand{\StringTok}[1]{\textcolor[rgb]{0.31,0.60,0.02}{#1}}

\providecommand{\tightlist}{%
  \setlength{\itemsep}{0pt}\setlength{\parskip}{0pt}}

\newlength{\cslhangindent}
\newlength{\csllabelwidth}
\newlength{\cslentryspacingunit} 
  {}%
  {\par}
\newenvironment{CSLReferences}[2] 
 {
  \setlength{\parindent}{0pt}
  \ifodd #1
  \let\oldpar\par
  \def\par{\hangindent=\cslhangindent\oldpar}
  \fi
  \setlength{\parskip}{#2\cslentryspacingunit}
 }%
 {}
\usepackage{calc}

\usepackage{booktabs}
\usepackage{longtable}
\usepackage{array}
\usepackage{multirow}
\usepackage{wrapfig}
\usepackage{float}
\usepackage{colortbl}
\usepackage{pdflscape}
\usepackage{tabu}
\usepackage{threeparttable}
\usepackage{threeparttablex}
\usepackage[normalem]{ulem}
\usepackage{makecell}
\usepackage{xcolor}
\usepackage{amsmath}
\usepackage{xparse}
\begin{document}
		
	\noindent\fbox{%
		\parbox{\textwidth}{%
			A peer reviewed, substantially rewritten version of this preprint is \href{https://doi.org/10.1007/s11135-025-02266-9	}{available here} and freely accessible: \newline
			\hfill \break
			Settanni, E., Srai, J.S. It’s a long way to the top (if you wanna biplot): a back-to-basics perspective on the implementation of principal component biplots in R. Qual Quant (2025). https://doi.org/10.1007/s11135-025-02266-9 \newline	 
			\hfill \break
			Please refer to the published version.
		}%
	}
	\newpage
	
\maketitle

\begin{abstract}
Principal Component Analysis and biplots are so well-established and
readily implemented that it is just too tempting to give for granted
their internal workings. In this note I get back to basics in comparing
how PCA and biplots are implemented in base-R and contributed R
packages, leveraging an implementation-agnostic understanding of the
computational structure of each technique. I do so with a view to
illustrating discrepancies that users might find elusive, as these arise
from seemingly innocuous computational choices made under the hood. The
proposed evaluation grid elevates aspects that are usually disregarded,
including relationships that should hold if the computational rationale
underpinning each technique is followed correctly. Strikingly, what is
expected from these equivalences rarely follows without caveats from the
output of specific implementations alone.
\end{abstract}


\hypertarget{introduction}{%
\section{Introduction}\label{introduction}}

A food scientist, a bioinformaticist, and a survey researcher walk into
a bar\ldots{} Jokes aside, what could they possibly have in common? The
answer is probably PCA---Principal Component Analysis.

Many will find the concept familiar, perhaps even trite. Overviews of
this technique leave little room for doubt about its popularity and
relevance across disciplines---see e.g., Knox
(\protect\hyperlink{ref-KnoxStevenW.2018Ml:t}{2018}, Ch. 11); Bro and
Smilde (\protect\hyperlink{ref-chemometric2014}{2014}); Abdi and
Williams (\protect\hyperlink{ref-AbdiWilliams}{2010}). Often, PCA is
complemented by a biplot---a closely related, yet distinct technique for
jointly visualizing observations and variables that typically make up a
data matrix (\protect\hyperlink{ref-DuToit1986GEDA}{du Toit, Steyn, and
Stumpf 1986} Ch. 6; \protect\hyperlink{ref-Gower2011}{Gower, Lubbe, and
LeRoux 2011}).

There is no shortage of options at the analysts' fingertip when
performing PCA: Tab.\ref{tab:tab1pdf} provides a non-exhaustive list.
Base-\texttt{R} alone provides two built-in functions besides several
contributed packages' own implementations. For a hands-on summary of
some of these functions see e.g., Mayor
(\protect\hyperlink{ref-MayorEric2015Lpaw}{2015}, Ch. 6); Kumar and Paul
(\protect\hyperlink{ref-Kumar2016Mtmw}{2016}, Ch. 4). Several references
on multivariate statistics cover both PCA and biplots in the context of
base \texttt{R} (e.g.,
\protect\hyperlink{ref-Venables.2002Masw}{Venables and Ripley 2002}, Ch.
11; \protect\hyperlink{ref-EverittBrian2011Aita}{Everitt and Hothorn
2011}, Ch. 3) or dedicated \texttt{R} packages (e.g.,
\protect\hyperlink{ref-Gower2011}{Gower, Lubbe, and LeRoux 2011}, Ch. 3;
\protect\hyperlink{ref-Pages2014MFAb}{Pagès 2014}, Ch. 1).

\begin{table}
\centering
\caption{\label{tab:tab1pdf}Selected implementations of PCA and biplots}
\centering
\fontsize{7}{9}\selectfont
\begin{tabular}[t]{lll}
\toprule
\multicolumn{1}{l}{ } & \multicolumn{2}{l}{Functions included} \\
\cmidrule(l{3pt}r{3pt}){2-3}
  & PCA & Biplots\\
\midrule
base-$\texttt{R}$ & $\texttt{prcomp()}$, $\texttt{princomp()}$ & $\texttt{biplot()}$\\
$\texttt{ade4}$ & $\texttt{dudi.pca()}$ & $\texttt{scatter()}$\\
$\texttt{amap}$ & $\texttt{acp()}$ & $\texttt{plot()}$\\
$\texttt{FactoMineR}$ & $\texttt{PCA()}$ & $\texttt{plot.PCA()}$\\
$\texttt{pcaMethods}$ & $\texttt{pca()}$ & $\texttt{slplot()}$\\
\addlinespace
$\texttt{PCAmixdata}$ & $\texttt{PCAmix()}$ & $\texttt{plot.PCAmix()}$\\
$\texttt{PCAtools}$ & $\texttt{pca()}$ & $\texttt{biplot()}$\\
$\texttt{psych}$ & $\texttt{principal()}$ & $\texttt{biplot.psych()}$\\
$\texttt{factoextra}$ &  & $\texttt{fviz\_pca\_biplot()}$\\
$\texttt{ggbiplot}$ &  & $\texttt{ggbiplot()}$\\
\bottomrule
\end{tabular}
\end{table}

With PCA and biplots being so readily implemented it is tempting to give
for granted their internal workings and go about it mechanically through
a canned routine of choice. The exception that proves the rule is the
package
\emph{\href{https://CRAN.R-project.org/package=LearnPCA}{LearnPCA}}
(\protect\hyperlink{ref-LearnPCA}{Hanson and Harvey 2022}). Its rich set
of vignettes is unique in its intent to address the self-directed
learner with a view to unpicking methodological aspects of PCA (but not
biplots) that \texttt{R} users rarely engage with, and whose importance
may be underplayed. It also provides comparative insights into
alternative ways of implementing PCA in base \texttt{R}.

Against this backdrop, what seems to be missing is---to the best of my
knowledge---a comparison of how PCA and biplots are implemented in base
\texttt{R} and contributed \texttt{R} packages, and how that compares
with an implementation-agnostic understanding of the computational
structure of each technique. In this note I attempt such comparison with
a view to illustrating discrepancies that users might find elusive, as
these arise from seemingly innocuous computational choices made under
the hood. By getting back to basics in PCA and biplots I elevate aspects
that are usually disregarded.

The remainder is structured as follows. The next section outlines an
implementation-agnostic understanding of the computational building
blocks in PCA and biplots with the aid of an illustrative example. Using
these insights as an evaluation grid, selected implementations are then
compared, pinpointing possible points of departure from what is
reasonably expected. A closing section summarises key practical
implications of these findings.

\hypertarget{buildblocks}{%
\section{Implementation-agnostic building blocks}\label{buildblocks}}

The seasoned practitioner might scoff at the idea of yet another
overview on PCA and swiftly move past. At the risk of disappointing
advanced readers, this section makes a point of reviewing the
computational building blocks of PCA and biplots---especially how they
come about---as an agnostic stance for comparing the output of specific
implementations.

A useful place to start is the visual intuition behind how PCA
works---i.e., that a set of data-points can be represented in a
lower-dimensional space while preserving relevant information about
them. The analogy with image compression often comes to mind
(\protect\hyperlink{ref-PooleDavid2015}{Poole 2014, 607};
\protect\hyperlink{ref-Peng_PCA}{Peng 2020}, Ch. 3). In the simplest
case, the data-points are 2-dimensional and the task at hand is to
determine their orthogonal projections onto an appropriately defined
line, thus obtaining a 1-dimensional representation of those points.
While simplistic, the pedagogic merits of a 2-dimensional visual example
are emphasised by the popularity of on-line resources such as Starmer
(\protect\hyperlink{ref-StatQuestPCA}{2018}) and amoeba
(\protect\hyperlink{ref-stackexchangePCA}{2015}). This is the case I
will consider throughout this section, by means of an illustrative
numerical example. Similar examples can be found elsewhere e.g., Hanson
and Harvey (\protect\hyperlink{ref-LearnPCA}{2022}); Abdi and Williams
(\protect\hyperlink{ref-AbdiWilliams}{2010}). The example is restricted
to numerical features, or variables, as PCA does not immediately apply
to categorical or binary features---see Kassambara and Mundt
(\protect\hyperlink{ref-PCAalternatives}{2020})'s taxonomy in the
context of package
\emph{\href{https://CRAN.R-project.org/package=factoextra}{factoextra}}.

\hypertarget{a-motivating-example}{%
\subsection{A motivating example}\label{a-motivating-example}}

\begin{figure}

{\centering \includegraphics[width=0.47\linewidth]{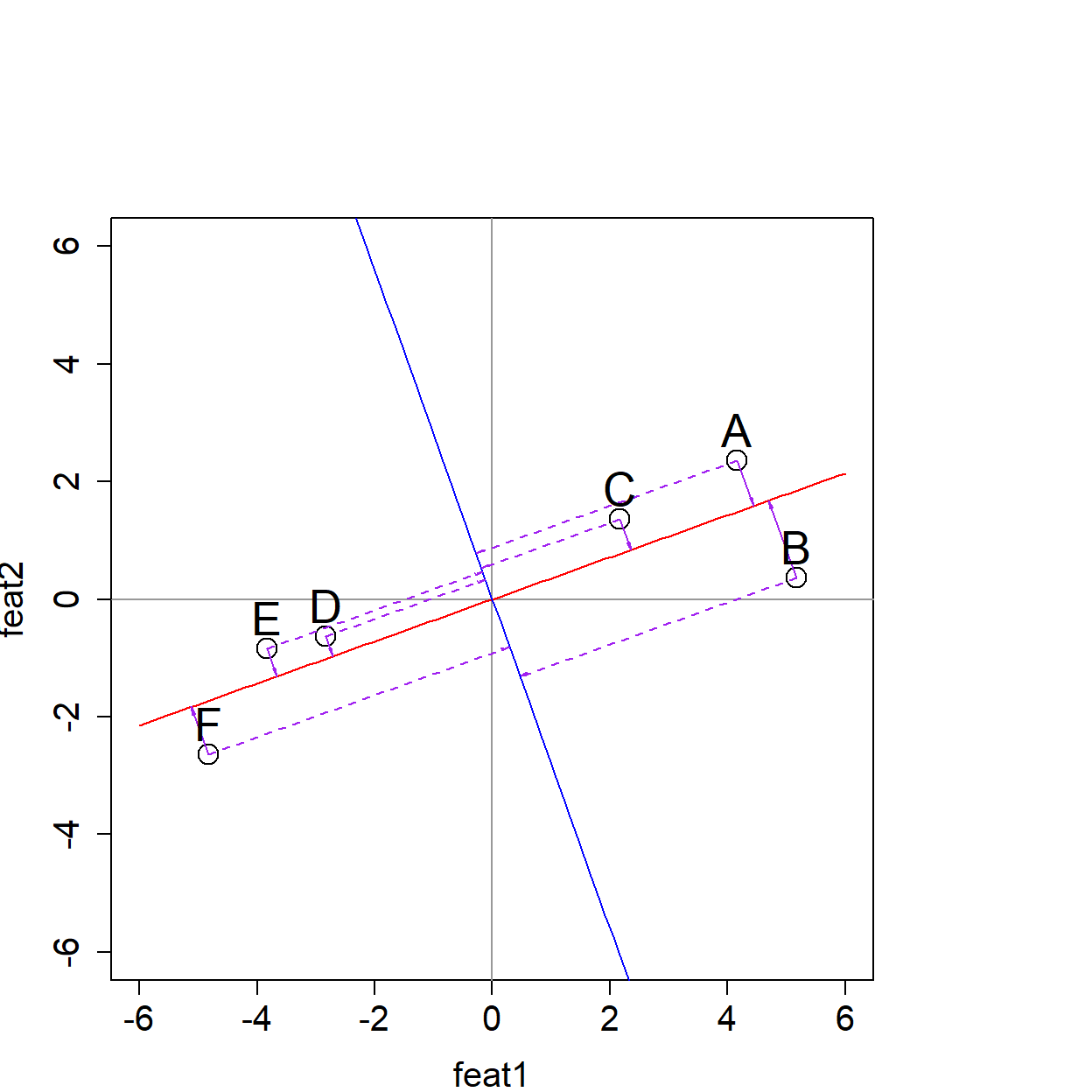} \includegraphics[width=0.47\linewidth]{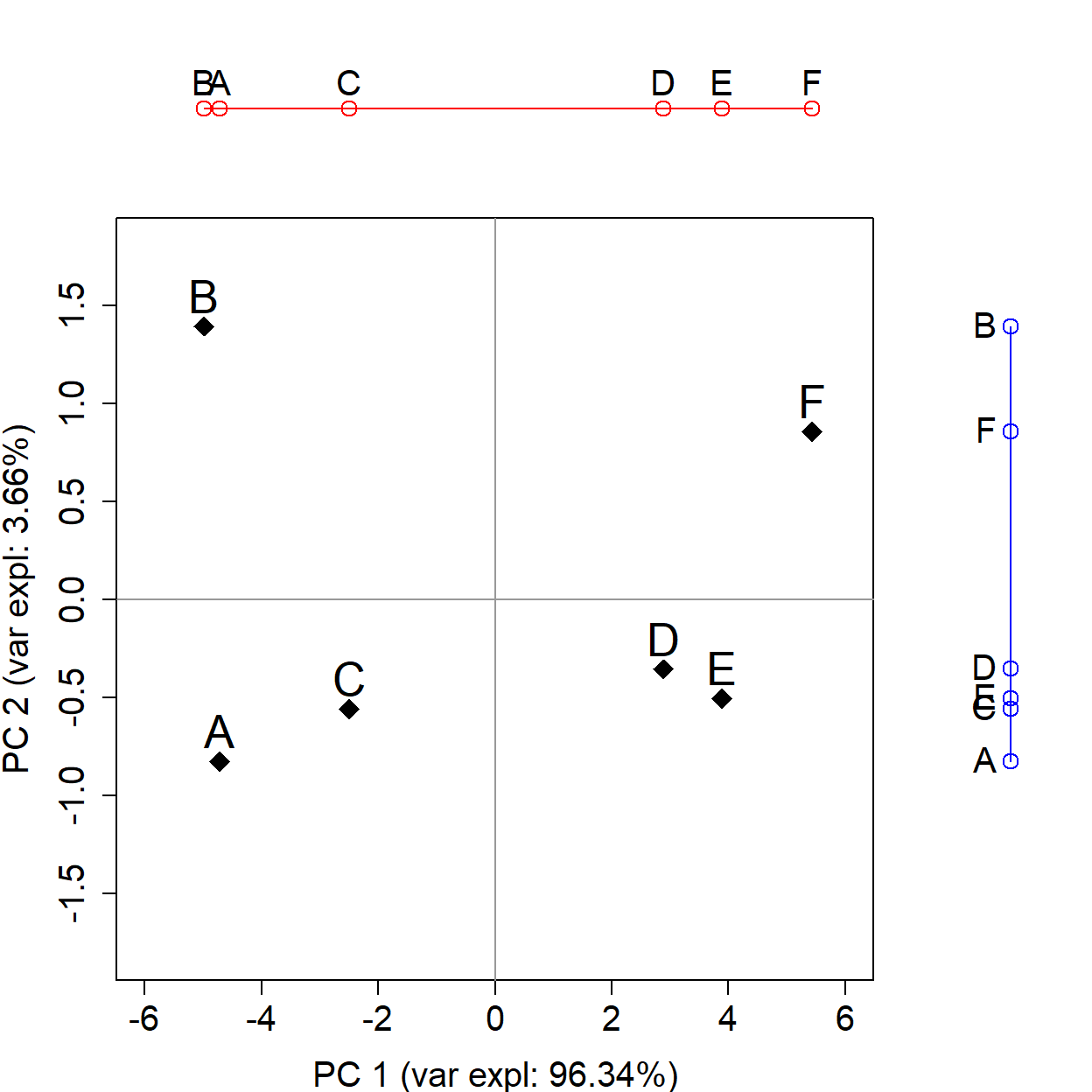} 

}

\caption{Visualisation of selected information from Tab.$\ref{tab:tab2pdf}$. Left: scatterplot of centred data-points corresponding to $\mathbf{Y}$ with orthogonal projections on two principal components represented by score vectors $\mathbf{z}_{\bullet1}$ (red) and $\mathbf{z}_{\bullet2}$ (blue); purple arrows represent distances bewteen points and porjection. Right: plot of observations using principal component scores as transformed coordinates.}\label{fig:fig1pdf}
\end{figure}

Consider the example in Tab.\ref{tab:tab2pdf}, and
Fig.\ref{fig:fig1pdf}. The first two columns of Tab.\ref{tab:tab2pdf}
represent the raw data, which are arranged in a so-called data matrix
\(\mathbf{X} = \lbrack x_{ij} \rbrack_{n \times m}\) consisting of
\(i=1,...,n\) observations and \(j=1,...,m\) features, or variables. In
the example \(n=6\) and \(m=2\). For each feature \(j\) the last two
rows of Tab.\ref{tab:tab2pdf} give its mean
\(\bar{x}_j=\frac{1}{n}\sum_{i}x_{ij}\), and sample variance
\(s_{x_j}^{2}=\frac{1}{n-1}\sum_i (x_{ij} - \bar{x}_j)^2\). In
\texttt{R} this is just: \texttt{apply(X,\ 2,\ mean)} and
\texttt{apply(X,\ 2,\ var)}, respectively.

\begin{table}

\caption{\label{tab:tab2pdf}An illustrative example with $m=2$ features and a $n=6$ observations}
\fontsize{7}{9}\selectfont
\begin{tabular}[t]{lrrrrrrrrrrrr}
\toprule
\multicolumn{1}{c}{ } & \multicolumn{4}{c}{Features} & \multicolumn{4}{c}{Optimal projections, PC1} & \multicolumn{4}{c}{Optimal projections, PC2} \\
\cmidrule(l{3pt}r{3pt}){2-5} \cmidrule(l{3pt}r{3pt}){6-9} \cmidrule(l{3pt}r{3pt}){10-13}
\multicolumn{1}{c}{ } & \multicolumn{2}{c}{Raw $\left[\mathbf{X}\right]$} & \multicolumn{2}{c}{Centred $\left[\mathbf{Y}\right]$} & \multicolumn{2}{c}{Proj. coord} & \multicolumn{2}{c}{Dist of proj. from} & \multicolumn{2}{c}{Proj. coord} & \multicolumn{2}{c}{Dist of proj. from} \\
\cmidrule(l{3pt}r{3pt}){2-3} \cmidrule(l{3pt}r{3pt}){4-5} \cmidrule(l{3pt}r{3pt}){6-7} \cmidrule(l{3pt}r{3pt}){8-9} \cmidrule(l{3pt}r{3pt}){10-11} \cmidrule(l{3pt}r{3pt}){12-13}
  & feat1 & feat2 & feat1 & feat2 & feat1 & feat2 & origin $\left[\mathbf{z}_{\bullet1}\right]$ & observ. & feat1 & feat2 & origin $\left[\mathbf{z}_{\bullet2}\right]$ & observ.\\
\midrule
A & 10.00 & 6.00 & 4.17 & 2.37 & 4.44 & 1.59 & -4.72 & 0.83 & -0.28 & 0.78 & -0.83 & 4.72\\
B & 11.00 & 4.00 & 5.17 & 0.37 & 4.70 & 1.68 & -4.99 & 1.39 & 0.47 & -1.31 & 1.39 & 4.99\\
C & 8.00 & 5.00 & 2.17 & 1.37 & 2.35 & 0.84 & -2.50 & 0.56 & -0.19 & 0.53 & -0.56 & 2.50\\
D & 3.00 & 3.00 & -2.83 & -0.63 & -2.71 & -0.97 & 2.88 & 0.36 & -0.12 & 0.34 & -0.36 & 2.88\\
E & 2.00 & 2.80 & -3.83 & -0.83 & -3.66 & -1.31 & 3.89 & 0.50 & -0.17 & 0.48 & -0.50 & 3.89\\
\addlinespace
F & 1.00 & 1.00 & -4.83 & -2.63 & -5.12 & -1.83 & 5.44 & 0.85 & 0.29 & -0.80 & 0.85 & 5.44\\
\midrule
mean & 5.83 & 3.63 & 0.00 & 0.00 & 0.00 & 0.00 & 0.00 & 0.75 & 0.00 & 0.00 & 0.00 & 4.07\\
sample var. & 18.97 & 3.13 & 18.97 & 3.13 & 18.88 & 2.41 & 21.28 & 0.14 & 0.09 & 0.72 & 0.81 & 1.41\\
\bottomrule
\end{tabular}
\end{table}

The columns of Tab.\ref{tab:tab2pdf} denoted as \(\mathbf{Y}\)
correspond to the centred data matrix, which is also of size
\(n \times m\), with generic element \(y_{ij} = x_{ij} - \bar{x}_j\).
The operation of centring \(\mathbf{X}\) is often expressed in matrix
notation as
\(\mathbf{Y}=\mathbf{X}-\frac{1}{n}\mathbf{1}\mathbf{1}^T\mathbf{X}\),
where \(\mathbf{1}\) is a unit vector of appropriate dimension
(\protect\hyperlink{ref-Venables.2002Masw}{Venables and Ripley 2002,
302}; \protect\hyperlink{ref-Gower2011}{Gower, Lubbe, and LeRoux 2011}).
In \texttt{R} this is accomplished with:
\texttt{Y\ \textless{}-\ apply(X,2,function(x)\ (x\ -\ mean(x)))}. The
mean and sample variance for the centred data are also shown in the last
two rows of Tab.\ref{tab:tab2pdf}. For each feature \(j\) the mean is
characteristically zero:
\(\bar{y}_j=\frac{1}{n}\sum_{i}y_{ij}=\frac{1}{n}\sum_{i}x_{ij}-\bar{x}_j=0\).
Yet the sample variance is the same as the raw data's:
\(s_{y_j}^{2}=\frac{1}{n-1}\sum_i (y_{ij} - \bar{y}_j)^2=\frac{1}{n-1}\sum_i (x_{ij} - \bar{x}_j - 0)^2=s_{x_j}\).

Contrary to what many think, so long as the quantities represented in
\(\mathbf{X}\) are expressed in the same units, or if they scale well
e.g., through a logarithmic transformation, it is not compulsory to
manipulate the data further (see e.g.,
\protect\hyperlink{ref-Venables.2002Masw}{Venables and Ripley 2002}).

Having centred the data, each row of \(\mathbf{Y}\) can be interpreted
as the coordinates of a point in the Cartesian plane corresponding to an
observation (row header). The \(n=6\) observations can then be plotted
as shown in Fig.\ref{fig:fig1pdf}, left-hand side. The same figure also
shows two lines passing through the origin---what is commonly referred
to as \emph{principal components}---and the orthogonal projections of
the centred data-points' onto each. For a given point, its projection is
found from intersecting a component, say the red line, and its
perpendicular passing through that point (see e.g.,
\protect\hyperlink{ref-intermedAlgebra}{Marecek 2017}, Ch. 3). The
position of an orthogonal projection onto a principal component, being
1-dimensional, is just its distance from the origin---or PCA
\emph{score}. The sign of a score, however, depends on which quadrant in
which the projection falls with respect to the origin. For the two
principal components in our example, the respective \emph{scores} are
denoted in Tab.\ref{tab:tab2pdf} by vectors \(\mathbf{z}_{\bullet1}\)
(red line) and \(\mathbf{z}_{\bullet2}\) (blue line).

What we see at work in Fig.\ref{fig:fig1pdf} is a key tenet of PCA,
albeit streamlined: a set of 2-dimensional data-points is transformed
into a set of 1-dimensional points, which are appropriately arranged
along a line, or principal component. In the context of PCA a set of
data-points is ``appropriately'' represented in a lower-dimensional
space when most information about its variance is retained. In our
example, the more dispersed the projected data-points along a
1-dimensional line, the better.

But how are the lines and projections in Fig.\ref{fig:fig1pdf} arrived
at, so that one can claim they are optimal? This aspect will be
discussed next, and will inform an implementation-agnostic evaluation
grid.

\hypertarget{PCAsection}{%
\subsection{Computational aspects of PCA}\label{PCAsection}}

To understand how the data-points' projections are positioned along a
1-dimensional \emph{principal component}, one can imagine rotating the
lines shown in Fig.\ref{fig:fig1pdf} (left) until the variance of the
projections' distances from the origin---i.e., the variance of the
\emph{scores} in \(\mathbf{z}_{\bullet1}\) and
\(\mathbf{z}_{\bullet1}\)---is maximal. For the first principal
component in our example, or PC1 (blue line), the last row in
Tab.\ref{tab:tab2pdf} shows that such variance is 21.3. This equals the
sum of the variances along the projections' coordinates, see columns 5
and 6 of Tab.\ref{tab:tab2pdf}. Typically, the variance of a principal
component's scores is expressed as a fraction of the total variance
along the the columns of \(\mathbf{Y}\). For PC1 in our example, the
maximal projections' variance achieved accounts for about 96\% of the
projected points'.

A well-established approach to obtain the optimal \emph{scores} is to
seek appropriate \emph{linear combinations} of the original data-points'
coordinates (e.g., \protect\hyperlink{ref-JolliffeI.T.2004Pca}{Jolliffe
2004}, Ch. 1 \& 3; \protect\hyperlink{ref-DuToit1986GEDA}{du Toit,
Steyn, and Stumpf 1986}, Ch. 9). In our example, the coordinates are
represented by the columns of the centred data matrix \(\mathbf{Y}\),
and a linear combination of interest is the matrix-vector product:

\begin{equation} \label{eq:lincomb}
  \mathbf{z} =
  \mathbf{Y}\mathbf{a} = \begin{bmatrix} a_1 y_{11} + a_2 y_{12} = z_1\\ 
    \vdots \\
    a_1 y_{i1} + a_2 y_{i2} = z_i \\
    \vdots \\
    a_1 y_{n1} + a_2 y_{n2} = z_n  \end{bmatrix}
\end{equation}

where \(\mathbf{a}=\begin{bmatrix} a_1, a_2 \end{bmatrix}^T\) is the
vector of unknown linear combination weights, commonly referred to as
\emph{loadings}; and \(\mathbf{z}\) is a vector of unknown principal
component \emph{scores} that position the projections of the data-points
in \(\mathbf{Y}\) along an optimally defined 1-dimensional line or
\emph{principal component}, as discussed.

The process of finding \(\mathbf{z}\) revolves around the sample
variance of its elements. The data being centred, \(\mathbf{z}\) has
mean \(\bar{z}=\frac{1}{n}\sum_{i}z_{i}=0\), from which it follows that:

\begin{align} \label{eq:var}
     \textrm{Var} \left(\mathbf{z} \right) = \quad & \frac{1}{n-1}\sum_i \left(z_i - \bar{z}\right)^2 \nonumber \\
         & = \frac{1}{n-1} \left(\mathbf{Ya} - \bar{z}\right)^T \cdot \left(\mathbf{Ya} - \bar{z} \right) \nonumber  \\
         & = \frac{1}{n-1}  \mathbf{a}^T\mathbf{Y}^T \mathbf{Ya} \nonumber \\
         & = \mathbf{a}^T\mathbf{S}\mathbf{a}
 \end{align}

where \(\mathbf{S}=\frac{1}{n-1}\mathbf{Y}^T\mathbf{Y}\) is the sample
covariance matrix of \(\mathbf{Y}\). In \texttt{R} this is simply:
\texttt{S\ \textless{}-\ cov(Y)}.

Based on the above, one must choose \(\mathbf{a}\) to maximize
\(\textrm{Var}\left( \mathbf{z} \right)\) while respecting orthogonality
constraints (\protect\hyperlink{ref-JolliffeI.T.2004Pca}{Jolliffe 2004,
4}; \protect\hyperlink{ref-Venables.2002Masw}{Venables and Ripley 2002,
303}; \protect\hyperlink{ref-EverittBrian2011Aita}{Everitt and Hothorn
2011}, Ch. 3). Since. from Equation \ref{eq:var},
\(\textrm{Var}\left( \mathbf{z} \right) = \mathbf{a}^T\mathbf{S}\mathbf{a}\),
this is typically framed as a mathematical program:

\begin{align}  \label{eq:program}
    \max_{\mathbf{a}} \quad & \mathbf{a}^T\mathbf{S}\mathbf{a} \nonumber \\
    \textrm{s.t.} \quad & \mathbf{a}^T\mathbf{a}=1
  \end{align}

As it turns out, solving the program in Equation \ref{eq:program} for
\(\mathbf{a}\) is equivalent to finding the principal eigenvector of
\(\mathbf{S}\). To see how that is so, I examine the first-order
conditions for a constrained maximum using Lagrange multipliers
(\protect\hyperlink{ref-JolliffeI.T.2004Pca}{Jolliffe 2004, 4}):

\begin{align} \label{eq:eigprob}
    \frac{\textrm{d} \mathbf{a}^T\mathbf{S}\mathbf{a} }{\textrm{d} \mathbf{a}} - \frac{\textrm{d}}{\textrm{d} \mathbf{a}} \lambda \left( \mathbf{a}^T\mathbf{a} -1\right) = \quad & 0 \nonumber \\
       2\mathbf{a}^T\mathbf{S} - 2\lambda \mathbf{a}^T  = \quad & 0 \nonumber \\
       \left( \mathbf{S} - \lambda \mathbf{I}\right)\mathbf{a} = \quad & 0 
\end{align}

where the vector derivative
\(\frac{\textrm{d} \mathbf{a}^T\mathbf{S}\mathbf{a} }{\textrm{d} \mathbf{a}}=2\mathbf{a}^T\mathbf{S}\)
is a special result of the general rule
\(\frac{\textrm{d} \mathbf{a}^T\mathbf{S}\mathbf{a}}{\textrm{d} \mathbf{a}} = \mathbf{a}^T\left(\mathbf{S}^T + \mathbf{S}\right)\)
when \(\mathbf{S}\) is symmetric
(\protect\hyperlink{ref-BinmoreDavies}{Binmore and Davies 2001, 162}),
which is the case for covariance matrices. The form of Equation
\ref{eq:eigprob} is typical of an eigenvalue problem in linear algebra
(e.g. \protect\hyperlink{ref-PooleDavid2015}{Poole 2014}, Ch. 4), and
promptly solved in \texttt{R} using the function \texttt{eigen(S)},
which in our example yields the eigenvalues of \(\mathbf{S}\)

\begin{Shaded}
\begin{Highlighting}[]
\FunctionTok{eigen}\NormalTok{(S)}\SpecialCharTok{$}\NormalTok{values}
\end{Highlighting}
\end{Shaded}

\begin{verbatim}
## [1] 21.28  0.81
\end{verbatim}

denoted as the vector
\(\boldsymbol{\lambda}=\begin{bmatrix} \lambda_1 & \lambda_{2} \end{bmatrix}^T\);
and the corresponding eigenvectors, juxtaposed

\begin{Shaded}
\begin{Highlighting}[]
\FunctionTok{eigen}\NormalTok{(S)}\SpecialCharTok{$}\NormalTok{vectors}
\end{Highlighting}
\end{Shaded}

\begin{verbatim}
##       [,1]  [,2]
## [1,] -0.94  0.34
## [2,] -0.34 -0.94
\end{verbatim}

denoted by the matrix
\(\mathbf{V} = \begin{bmatrix} \mathbf{a}_{\bullet1} & \mathbf{a}_{\bullet2} \end{bmatrix}\).
The above suggests there are two linear combinations---i.e., as many as
there are features---that could satisfy Equation \ref{eq:lincomb}.
Jolliffe (\protect\hyperlink{ref-JolliffeI.T.2004Pca}{2004, 5})
demonstrates that the same reasoning leading to Equation
\ref{eq:eigprob} applies recursively beyond the first \emph{principal
component}.

If one considers only the linear combination associated with the largest
eigenvalue \(\lambda = \lambda_1 =21.28\) and the corresponding
eigenvector
\(\mathbf{a} = \mathbf{a}_{\bullet1}=\begin{bmatrix} -0.94, -0.34 \end{bmatrix}^T\)
of \(\mathbf{S}\), Equation \ref{eq:lincomb} yields the \emph{scores}
vector \(\mathbf{z}=\mathbf{z}_{\bullet1}\) shown in Tab.
\ref{tab:tab2pdf}. Choosing
\(\mathbf{a} = \mathbf{a}_{\bullet2} =\begin{bmatrix} 0.34, -0.94 \end{bmatrix}^T\)
yields \(\mathbf{z}=\mathbf{z}_{\bullet2}\), instead.

These vectors may be juxtaposed to form a \emph{scores} matrix
\(\mathbf{Z} = \begin{bmatrix} \mathbf{z}_{\bullet ,1} & \mathbf{z}_{\bullet,2} \end{bmatrix} = \mathbf{YV}\).
Each column of the \emph{scores} matrix \(\mathbf{Z}\) is a
lower-dimensional representation from which one can reconstruct the
2-dimensional data-points in \(\mathbf{Y}\)
(\protect\hyperlink{ref-Gower2011}{Gower, Lubbe, and LeRoux 2011}). For
example, columns 5 and 6 in Tab.\ref{tab:tab2pdf} are reconstructed as
\(\mathbf{Y}_{\textrm{PC1}} = \mathbf{z}_{\bullet1} \mathbf{a}_{\bullet1}^T\).
In \texttt{R} this is:
\texttt{outer(Z{[},1{]},\ eigen(S)\$vectors{[},1{]})}.

Combining Equations \ref{eq:eigprob} and \ref{eq:var} brings about an
often overlooked equivalence between the variance along a given
\emph{scores} vector \(\mathbf{z}\) and the corresponding eigenvalue
\(\lambda\) of \(\mathbf{S}\):

\begin{align} \label{eq:eigvar}
  \textrm{Var}\left[\mathbf{z}\right] \quad & =  \mathbf{a}^T\mathbf{S}\mathbf{a} \nonumber \\   
  \quad & =  \mathbf{a}^T\lambda\mathbf{a} \nonumber \\   
  \quad & = \lambda 
\end{align}

For example, assuming \(\mathbf{z}=\mathbf{z}_{\bullet1}\) one verifies
in \texttt{R} that:

\begin{Shaded}
\begin{Highlighting}[]
\NormalTok{Z }\OtherTok{\textless{}{-}}\NormalTok{ Y }\SpecialCharTok{\%*\%} \FunctionTok{eigen}\NormalTok{(S)}\SpecialCharTok{$}\NormalTok{vectors}
\FunctionTok{all.equal}\NormalTok{(}\FunctionTok{apply}\NormalTok{(Z,}\DecValTok{2}\NormalTok{,var)[}\DecValTok{1}\NormalTok{], }\FunctionTok{eigen}\NormalTok{(S)}\SpecialCharTok{$}\NormalTok{values[}\DecValTok{1}\NormalTok{])}
\end{Highlighting}
\end{Shaded}

\begin{verbatim}
## [1] TRUE
\end{verbatim}

Although it is simplistic to assume two features, the first two
principal components have well-established uses for visualising the
outputs of PCA, even when there are more than two features.
Traditionally, a 2-dimensional visualisation of PCA is attained by
combining the first two principal components in a plot as shown in
Fig.\ref{fig:fig1pdf} (right), where the vectors
\(\mathbf{z}_{\bullet,1}\) and \(\mathbf{z}_{\bullet,2}\) provide the
``transformed'' system of coordinates and the original \(n=6\)
observations are plotted accordingly. In the presence of more than two
features one would typically retain the first two components with the
caveat that the quality of such an approximation is given by the
relative weight of the two largest eigenvalues of the covariance matrix
\(\mathbf{S}\). From Equation \ref{eq:eigvar}, this is equivalent to the
proportion of variance associated with the first two principal
components relative to the variance of the projected data-points.

\hypertarget{SVDsection}{%
\subsection{Alternative approaches: Singular Value
Decomposition}\label{SVDsection}}

The computational building blocks described so far emphasise the logical
path from seeking a variance-maximising linear combination in Equation
\ref{eq:lincomb} to solving the eigenvalue problem in Equation
\ref{eq:eigprob}. In practice, the preferred way to obtain
\emph{loadings} and \emph{scores} is by doing a Singular Value
Decomposition (SVD) of the rectangular matrix \(\mathbf{Y}\) (e.g.,
\protect\hyperlink{ref-JolliffeI.T.2004Pca}{Jolliffe 2004}, Ch. 3;
\protect\hyperlink{ref-LearnPCA}{Hanson and Harvey 2022}):

\begin{equation} \label{eq:svdbasic}
  \mathbf{Y}=\mathbf{UD}\mathbf{V}^T
\end{equation}

which is equivalent to the command \texttt{svd(Y)} in \texttt{R}. With
reference to our simple example, Equation \ref{eq:svdbasic} generates
the following standard outputs
(\protect\hyperlink{ref-PooleDavid2015}{Poole 2014}, Ch. 7):

\begin{itemize}
\item
  A matrix
  \(\mathbf{U}=\begin{bmatrix}\mathbf{u}_{\bullet,1} & \mathbf{u}_{\bullet,3}\end{bmatrix}\)
  formed with the ``left'' eigenvectors of \(\mathbf{Y}\):

\begin{Shaded}
\begin{Highlighting}[]
 \FunctionTok{svd}\NormalTok{(Y)}\SpecialCharTok{$}\NormalTok{u}
\end{Highlighting}
\end{Shaded}

\begin{verbatim}
##       [,1]  [,2]
## [1,] -0.46  0.41
## [2,] -0.48 -0.69
## [3,] -0.24  0.28
## [4,]  0.28  0.18
## [5,]  0.38  0.25
## [6,]  0.53 -0.42
\end{verbatim}
\item
  A matrix
  \(\mathbf{V}=\begin{bmatrix}\mathbf{v}_{\bullet1} & \mathbf{v}_{\bullet2}\end{bmatrix}\)
  formed with the ``right'' eigenvectors of \(\mathbf{Y}\)

\begin{Shaded}
\begin{Highlighting}[]
\FunctionTok{svd}\NormalTok{(Y)}\SpecialCharTok{$}\NormalTok{v}
\end{Highlighting}
\end{Shaded}

\begin{verbatim}
##       [,1]  [,2]
## [1,] -0.94 -0.34
## [2,] -0.34  0.94
\end{verbatim}

  which is analogous to the juxtaposed \emph{loadings} vectors. Yet
  using \(\mathbf{V}\) to indicate both
  \(\begin{bmatrix} \mathbf{a}_{\bullet1} & \mathbf{a}_{\bullet2} \end{bmatrix}\)
  from Equation \ref{eq:eigprob} and
  \(\begin{bmatrix}\mathbf{v}_{\bullet1} & \mathbf{v}_{\bullet2}\end{bmatrix}\)
  is an abuse of notation, since the elements' signs may differ.
\item
  A vector
  \(\boldsymbol{\ell}=\begin{bmatrix} l_1 & l_2 \end{bmatrix}^T\) of
  ``singular values'' of \(\mathbf{Y}\), which is typically diagonalised
  to form the matrix
  \(\mathbf{D} = \begin{bmatrix} l_1 & 0 \\ 0 & l_2 \end{bmatrix}\) in
  Equation \ref{eq:svdbasic}:

\begin{Shaded}
\begin{Highlighting}[]
\FunctionTok{svd}\NormalTok{(Y)}\SpecialCharTok{$}\NormalTok{d}
\end{Highlighting}
\end{Shaded}

\begin{verbatim}
## [1] 10.32  2.01
\end{verbatim}
\end{itemize}

It is worth emphasising that the singular values in
\(\boldsymbol{\ell}\) are distinct from, but related to the eigenvalues
in \(\boldsymbol{\lambda}\) (e.g.,
\protect\hyperlink{ref-LearnPCA}{Hanson and Harvey 2022};
\protect\hyperlink{ref-JolliffeI.T.2004Pca}{Jolliffe 2004, 38}). In our
example, such relationship is:

\begin{align} \label{eq:svev}
       \boldsymbol{\ell} \quad & = (n-1)^{1/2} \begin{bmatrix} \lambda_1^{1/2} & \lambda_2^{1/2} \end{bmatrix}  \nonumber \\
                  \quad & = (n-1)^{1/2} \begin{bmatrix} \sqrt{\textrm{Var}\left[\mathbf{z}_{\bullet,1}\right]} & \sqrt{\textrm{Var}\left[\mathbf{z}_{\bullet,1}\right]} \end{bmatrix} \nonumber \\
                  \quad & = (n-1)^{1/2} \begin{bmatrix} \sigma_1 & \sigma_2 \end{bmatrix}
\end{align}

where \(\sigma_j = \textrm{s.d.}[\mathbf{z}_{\bullet j}]\); and
\(\lambda_j = \textrm{Var}[ \mathbf{z}_{\bullet j}]\), based on Equation
\ref{eq:eigvar}. One can verify in \texttt{R} that:

\begin{Shaded}
\begin{Highlighting}[]
\FunctionTok{all.equal}\NormalTok{(}\FunctionTok{sqrt}\NormalTok{((}\FunctionTok{nrow}\NormalTok{(Y)}\SpecialCharTok{{-}}\DecValTok{1}\NormalTok{)}\SpecialCharTok{*}\FunctionTok{eigen}\NormalTok{(S)}\SpecialCharTok{$}\NormalTok{values), }\FunctionTok{sqrt}\NormalTok{(}\FunctionTok{nrow}\NormalTok{(Y)}\SpecialCharTok{{-}}\DecValTok{1}\NormalTok{)}\SpecialCharTok{*}\FunctionTok{apply}\NormalTok{(Z,}\DecValTok{2}\NormalTok{,sd))}
\end{Highlighting}
\end{Shaded}

\begin{verbatim}
## [1] TRUE
\end{verbatim}

To obtain the \emph{scores} matrix \(\mathbf{Z}\) from an SVD, one
right-multiplies both sides of Equation \ref{eq:svdbasic} by
\(\mathbf{V}\):

\begin{align} \label{eq:svdscores}
  \mathbf{YV} & = \mathbf{UD}\mathbf{V}^T \mathbf{V} \nonumber \\
              & = \mathbf{UD} \nonumber \\
              & = \mathbf{Z}_{SVD} 
\end{align}

In practice, \(\mathbf{Z}_{SVD}\) and
\(\mathbf{Z} = \begin{bmatrix} \mathbf{z}_{\bullet ,1} & \mathbf{z}_{\bullet,2} \end{bmatrix}\),
which is shown in Tab.\ref{tab:tab2pdf}, are often used interchangeably.
Yet they are only equivalent in absolute values. In our example, one can
verify that:

\begin{Shaded}
\begin{Highlighting}[]
 \FunctionTok{all.equal}\NormalTok{(}\FunctionTok{abs}\NormalTok{(}\FunctionTok{unname}\NormalTok{(U }\SpecialCharTok{\%*\%}\NormalTok{ D)), }\FunctionTok{abs}\NormalTok{(}\FunctionTok{unname}\NormalTok{(Y }\SpecialCharTok{\%*\%} \FunctionTok{eigen}\NormalTok{(S)}\SpecialCharTok{$}\NormalTok{vector)))}
\end{Highlighting}
\end{Shaded}

\begin{verbatim}
## [1] TRUE
\end{verbatim}

\hypertarget{sectionBiplots}{%
\subsection{Principal components biplots}\label{sectionBiplots}}

The typical PCA plot in Fig.\ref{fig:fig1pdf} (right) may be enriched to
become a \emph{biplot}. The difference between these concepts might seem
cosmetic: both rely on a ``transformed'' system of coordinates based on
PCA \emph{scores}, but differ in scope as to what they seek to
visualise. The former only visualises observations i.e., the rows of the
data-matrix; whereas the latter aims to jointly represent observations
and features i.e., both the rows and the columns of the data-matrix
(\protect\hyperlink{ref-JolliffeI.T.2004Pca}{Jolliffe 2004}, Ch. 5).

On a deeper level, the task of attaining an overlaid representation of a
plot of features and a plot of observations in a \emph{biplot} is
underpinned by a distinct computational strategy. Whilst it was
straightforward to derive the plot in Fig.\ref{fig:fig1pdf} from
previous computations, a \emph{biplot} is inherently associated with an
SVD of the centred data-matrix (\protect\hyperlink{ref-Gower2011}{Gower,
Lubbe, and LeRoux 2011};
\protect\hyperlink{ref-Venables.2002Masw}{Venables and Ripley 2002}, Ch.
11; \protect\hyperlink{ref-DuToit1986GEDA}{du Toit, Steyn, and Stumpf
1986}, Ch. 6; \protect\hyperlink{ref-ggbiplot_vignette}{Vu and Friendly
2024}).

In the context of principal component biplots, the SVD in Equation
\ref{eq:svdbasic} serves as the starting point, but is modified so that
the diagonal matrix of singular values is ``split'' based on a parameter
\(0\le \alpha \le1\). Such modification is illustrated below in the
context of our simplified example, with just two features:

\begin{align} \label{eq:svdbiplot}
  \mathbf{Y}  & = \mathbf{UD}\mathbf{V}^T \nonumber \\
              & = \mathbf{U}\mathbf{L}^{\alpha}\mathbf{L}^{1-\alpha}\mathbf{V}^T \nonumber \\ 
              & = \begin{bmatrix} \mathbf{u}_{\bullet,1} & \mathbf{u}_{\bullet,2} \end{bmatrix} 
                           \begin{bmatrix} l_1 &  0 \\ 0 & l_2 \end{bmatrix}^{\alpha} 
                           \begin{bmatrix} l_1 &  0 \\ 0 & l_2 \end{bmatrix}^{1-\alpha}
                           \begin{bmatrix} \mathbf{v}_{\bullet,1}\\ \mathbf{v}_{\bullet,2} \end{bmatrix}^{\alpha} \nonumber \\
              & = \begin{bmatrix} u_{11}l_1^{\alpha} & u_{12}l_2^{\alpha}  \nonumber \\
                                           u_{21}l_1^{\alpha} & u_{22}l_2^{\alpha}  \nonumber \\ 
                                           \vdots & \vdots \nonumber \\
                                           u_{n1}l_1^{\alpha} & u_{n2}l_2^{\alpha}  \nonumber \\
                           \end{bmatrix}
                           \begin{bmatrix} v_{11}l_1^{(1-\alpha)} & v_{21}l_1^{(1-\alpha)} \nonumber \\
                                           v_{12}l_2^{(1-\alpha)} & v_{22}l_2^{(1-\alpha)}  \nonumber \\ 
                           \end{bmatrix}  \nonumber \\
              & = \mathbf{A}\mathbf{B}^T
\end{align}

In the presence of more than two features, the right-hand side of
Equation \ref{eq:svdbiplot} yields a rank-2 approximation of the centred
data-matrix i.e.,

\begin{equation} \label{eq:ranktwo}
   \mathbf{Y} \approx \mathbf{Y}_{(2)} = \mathbf{A}\mathbf{B}^T
\end{equation}

When \(\alpha=0\) Equation \ref{eq:svdbiplot} is referred to as a
\emph{principal component biplot}, which is of interest here. In our
example with just two features, the matrices \(\mathbf{A}\) and
\(\mathbf{B}\) take on the following meaning:

\begin{itemize}
\tightlist
\item
  The coordinates of the observations along the principal components are
  given by \(\mathbf{A}=\mathbf{U}\). Based on Equation
  \ref{eq:svdscores}, and using interchangeably \(\mathbf{Z}\) and
  \(\mathbf{Z}_{SVD}\) for simplicity, it can be shown that this is
  equivalent to scaling the PCA plot coordinates given by the
  \emph{scores} matrix:
\end{itemize}

\begin{align} \label{eq:obscoordbiplot}
 \mathbf{A} & = \mathbf{U} \nonumber \\
            & = \mathbf{Z}\mathbf{D}^{-1} \nonumber \\
            & =\begin{bmatrix}\mathbf{z}_{\bullet,1}l_1^{-1} & \mathbf{z}_{\bullet,2}l_2^{-1}\end{bmatrix}
\end{align}

\begin{itemize}
\tightlist
\item
  Simultaneously, the coordinates of the features are given by the
  matrix product \(\mathbf{B}^T=\mathbf{D}\mathbf{V}^T\) between the
  right eigenvectors and the singular values of an SVD. Yet this result
  can be re-expressed in terms of the \emph{loadings} and
  \emph{scores}'s St.Dev. Based on Equation \ref{eq:svev}:
\end{itemize}

\begin{align} \label{eq:pcasd}
   \mathbf{B}^T & = \mathbf{D}\mathbf{V}^T \nonumber \\
                & = \begin{bmatrix}\mathbf{v}_{\bullet,1} l_1 \\ \mathbf{v}_{\bullet,2} l_2\end{bmatrix} \nonumber \\
                & = (n-1)^{1/2}\begin{bmatrix}\mathbf{v}_{\bullet,1} \sigma_1 \\ 
                    \mathbf{v}_{\bullet,2}  \sigma_2 \end{bmatrix}
 \end{align}

In our simple example the matrix \(\mathbf{A}\) of observations
coordinates is equivalent to the matrix \(\mathbf{U}\) of left
eigenvectors previously obtained from Equation \ref{eq:svdbasic}, as
expected:

\begin{Shaded}
\begin{Highlighting}[]
\FunctionTok{all.equal}\NormalTok{(Z\_svd }\SpecialCharTok{\%*\%} \FunctionTok{diag}\NormalTok{(}\DecValTok{1}\SpecialCharTok{/}\FunctionTok{svd}\NormalTok{(Y)}\SpecialCharTok{$}\NormalTok{d), }\FunctionTok{svd}\NormalTok{(Y)}\SpecialCharTok{$}\NormalTok{u)}
\end{Highlighting}
\end{Shaded}

\begin{verbatim}
## [1] TRUE
\end{verbatim}

whereas the coordinates for the features given by matrix
\(\mathbf{B}^T\) are:

\begin{Shaded}
\begin{Highlighting}[]
\NormalTok{D }\SpecialCharTok{\%*\%} \FunctionTok{t}\NormalTok{(V)}
\end{Highlighting}
\end{Shaded}

\begin{verbatim}
##       [,1]  [,2]
## [1,] -9.71 -3.47
## [2,] -0.68  1.89
\end{verbatim}

\begin{figure}

{\centering \includegraphics[width=0.5\linewidth]{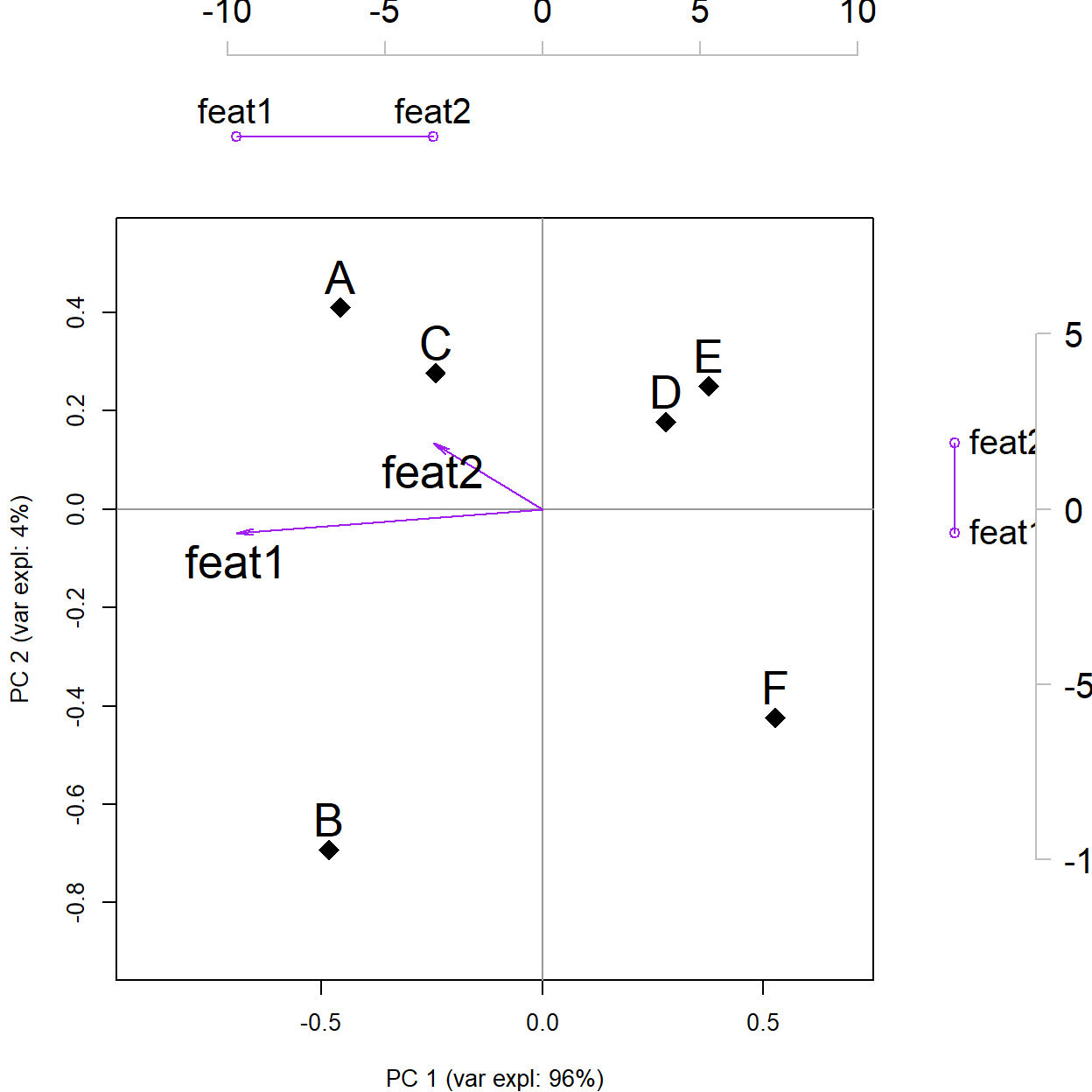} 

}

\caption{Biplot of observations and features for an illustrative example based on SVD. The superposed dual axes  (coloured grey) reflect the different scale of the features' coordinates.}\label{fig:fig2pdf}
\end{figure}

For our example, a joint representation of observations and features
based on \(\mathbf{A}\) and \(\mathbf{B}\),which are characteristic of a
principal components biplot, is shown in Fig.\ref{fig:fig2pdf}.

What is clear from Equations \ref{eq:obscoordbiplot} and \ref{eq:pcasd}
is that there is more to Fig.\ref{fig:fig2pdf} than just overlaying a
crude plot of \emph{loadings} over a plot of PCA \emph{scores}---which
happens frequently in practice. If the observations' and features'
coordinates are jointly computed as shown, one can verify certain
identities between the columns of matrix \(\mathbf{B}\) and the standard
deviation of the corresponding features, or their correlation. Such
properties are often overlooked, but useful for comparative purposes and
will be illustrated next.

\hypertarget{biplot-properties-linked-to-features}{%
\subsection{Biplot properties linked to
features}\label{biplot-properties-linked-to-features}}

The literature often states that, for a given feature \(j\) the length
of the vector formed by its coordinates on a biplot should be equivalent
to the feature's standard deviation (St.Dev.) \(\sigma_j\) (e.g.,
\protect\hyperlink{ref-Venables.2002Masw}{Venables and Ripley 2002,
312}; \protect\hyperlink{ref-DuToit1986GEDA}{du Toit, Steyn, and Stumpf
1986, 108}). This statement is hardly self-evident and might generate
some confusion. To demonstrate it, I follow Jolliffe
(\protect\hyperlink{ref-JolliffeI.T.2004Pca}{2004, 77}) and
left-multiply both sides of Equation \ref{eq:svdbiplot} by
\(\mathbf{Y}^T\):

\begin{align} \label{eq:vectorlen1}
  \mathbf{Y}^T\mathbf{Y} & = \mathbf{Y}^T\mathbf{A}\mathbf{B}^T \nonumber  \\
                  & = \mathbf{B}\left(\mathbf{U}\mathbf{L}^0 \right)^T\left(\mathbf{U}\mathbf{L}^0 \right)\mathbf{B}^T \nonumber \\
                  & = \mathbf{B}\mathbf{U}^T\mathbf{U}\mathbf{B}^T \nonumber \\
  (n-1)\mathbf{S} & = \mathbf{B}\mathbf{B}^T 
\end{align}

where \(\mathbf{A}=\mathbf{U}\mathbf{L}^\alpha\) from Equation
\ref{eq:svdbiplot}, and \(\alpha=0\) is characteristic of a principal
components biplot. Also, the columns of \(\mathbf{U}\) are orthonormal
hence \(\mathbf{U}^T\mathbf{U}=\mathbf{I}\). As before, \(\mathbf{S}\)
is the covariance matrix of \(\mathbf{Y}\).

Next I focus on a generic diagonal element \(s_{jj}\) of the covariance
matrix \(\mathbf{S}\) on the left-hand side of Equation
\ref{eq:vectorlen1} to show how the St.Dev. \(\sigma^{\textrm{feat}}_j\)
of the \(j\)-th feature relates to the length of vector
\(\mathbf{b}_{j \bullet}\), which corresponds to the \(j\)-th column of
matrix \(\mathbf{B}^T\) in Equation \ref{eq:pcasd}:

\begin{align} \label{eq:length}
     (n-1)s_{jj} \quad & = \mathbf{b}_{j \bullet}^T \mathbf{b}_{j \bullet} \nonumber \\
        \sigma^{\textrm{feat}}_j \quad & = (n-1)^{-\frac{1}{2}} \left\lVert \mathbf{b}_{j \bullet}  \right\rVert_2 \nonumber \\  
                 \quad & = \sqrt{\frac{1}{n-1}\sum_k\left( v_{jk}l_k\right)^2} \nonumber \\
                 \quad & = \sqrt{\sum_k \left(v_{jk} \sqrt{\lambda_k} \right)^2} 
\end{align}

where \(l_j = \left( {n-1} \right)^{1/2} \sqrt{\lambda_j}\) based on
Equation \ref{eq:svev}. For our example, one can verify in \texttt{R}
that:

\begin{Shaded}
\begin{Highlighting}[]
\FunctionTok{all.equal}\NormalTok{(}\FunctionTok{sqrt}\NormalTok{(}\FunctionTok{diag}\NormalTok{(S)), }\FunctionTok{apply}\NormalTok{(}\FunctionTok{t}\NormalTok{(B),}\DecValTok{2}\NormalTok{,}\ControlFlowTok{function}\NormalTok{(x)\{}\FunctionTok{norm}\NormalTok{(x,}\StringTok{"2"}\NormalTok{)\})}\SpecialCharTok{/}\FunctionTok{sqrt}\NormalTok{(}\FunctionTok{nrow}\NormalTok{(Y)}\SpecialCharTok{{-}}\DecValTok{1}\NormalTok{))}
\end{Highlighting}
\end{Shaded}

\begin{verbatim}
## [1] TRUE
\end{verbatim}

Confusions may arise due to the form of the right-hand side in Equation
\ref{eq:length}. For instance, it would be imprecise to conclude---as
the phrasing in some references might suggest---that the vector whose
length equals the St.Dev. of a given feature \(j\) is the corresponding
column of the matrix \(\mathbf{B}^T\) in Equation \ref{eq:pcasd}. Also,
it would be misleading to expect that Equation \ref{eq:length} holds for
a rank-2 approximation when there are more than two features, see
Equation \ref{eq:ranktwo}.

A second property associated with principal components biplots is that
the cosine similarity between a pair of vectors of features coordinates
is equivalent to the correlation coefficient between these features,
assuming these have been centred
(\protect\hyperlink{ref-JolliffeI.T.2004Pca}{Jolliffe 2004, 77};
\protect\hyperlink{ref-DuToit1986GEDA}{du Toit, Steyn, and Stumpf 1986,
108}). In our example, the presence of two features simplifies the task
of illustrating this point.

The two features vectors shown in Fig.\ref{fig:fig2pdf}
(purple-coloured) correspond to the columns of \(\mathbf{B}\) in
Equation \ref{eq:pcasd}. I denote these as \(\mathbf{b}_{\bullet1}\) and
\(\mathbf{b}_{\bullet2}\). Squaring the 2-norm of their difference
\(\left\lVert \mathbf{b}_{\bullet1} - \mathbf{b}_{\bullet2} \right\rVert_2^2\)
leads to the following ``textbook'' equivalence---for details see
Binmore and Davies (\protect\hyperlink{ref-BinmoreDavies}{2001, 18}):

\begin{align} \label{eq:cos}
  \cos\theta \quad & = \frac{\langle{\mathbf{b}_{\bullet1}, \mathbf{b}_{\bullet2}}\rangle}{\left\lVert \mathbf{b}_{\bullet1} \right\rVert_2 \cdot  \left\lVert \mathbf{b}_{\bullet2}  \right\rVert_2} 
\end{align}

which is the definition of cosine similarity. In Equation \ref{eq:cos}
the angle in radians between two features vectors is \(\theta\); and
\(\langle{\cdot,\cdot} \rangle\) denotes the inner product between them.
What is rarely shown is that Equation \ref{eq:cos} is equivalent to the
\emph{correlation} between the two features in our simple example. This
equivalence can be proved starting from the correlation coefficient
between two columns of matrix \(\mathbf{Y}\):

\begin{equation} \label{eq:corr}
     \textrm{corr}_{\mathbf{y}_{\bullet1},\mathbf{y}_{\bullet2}} = \frac{\langle{\mathbf{y}_{\bullet 1}, \mathbf{y}_{\bullet 2}}\rangle}{\left\lVert \mathbf{y}_{\bullet1}\right\rVert_2 \cdot \left\lVert \mathbf{y}_{\bullet2}\right\rVert_2} 
\end{equation}

Recalling Equations \ref{eq:obscoordbiplot} and \ref{eq:pcasd}, I denote
the \(i\)-th row of matrix \(\mathbf{A}\) as \(\mathbf{u}_{i\bullet}\)
and rephrase the inner product in Equation \ref{eq:corr} as:

\begin{align} \label{eq:buub}
       \langle{\mathbf{y}_{\bullet 1}, \mathbf{y}_{\bullet 2}}\rangle \quad & =   \langle{ \begin{bmatrix} 
       \mathbf{u}_{1\bullet} \cdot \mathbf{b}_{\bullet1}  &
       \dots &
       \mathbf{u}_{i\bullet} \cdot \mathbf{b}_{\bullet1} &
       \dots & 
       \mathbf{a}_{n\bullet} \cdot \mathbf{b}_{\bullet1}  \end{bmatrix},
       \begin{bmatrix}
            \mathbf{u}_{1\bullet} \cdot \mathbf{b}_{\bullet2} &
            \dots &
            \mathbf{u}_{i\bullet} \cdot \mathbf{b}_{\bullet2} &
            \dots &
            \mathbf{u}_{n\bullet} \cdot \mathbf{b}_{\bullet2} \end{bmatrix}}\rangle \nonumber \\
      \quad & =\left(\mathbf{U}\mathbf{b}_{\bullet1}\right)^T \mathbf{U}\mathbf{b}_{\bullet2}
\end{align}

The equivalence in Equation \ref{eq:buub} also applies to the
denominator of Equation \ref{eq:corr} since, for a given feature \(j\),
\(\left\lVert \mathbf{y}_{\bullet j}\right\rVert_2 = \sqrt{\langle{\mathbf{y}_{\bullet j}, \mathbf{y}_{\bullet j}}\rangle}\).
Hence, substituting Equation \ref{eq:buub} in Equation \ref{eq:corr}
yields:

\begin{align} \label{eq:coscorr}
  \frac{\langle{\mathbf{y}_{\bullet 1}, \mathbf{y}_{\bullet 2}}\rangle}{\left\lVert \mathbf{y}_{\bullet1}\right\rVert_2 \cdot \left\lVert \mathbf{y}_{\bullet2}\right\rVert_2} \quad & = \frac{\mathbf{b}_{\bullet1}^T\mathbf{U}^T \mathbf{U}\mathbf{b}_{\bullet2}}{\sqrt{\mathbf{b}_{\bullet 1}^T\mathbf{U}^T \mathbf{U}\mathbf{b}_{\bullet 1}}\sqrt{\mathbf{b}_{\bullet 2}^T\mathbf{U}^T \mathbf{U}\mathbf{b}_{\bullet 2}}} \nonumber \\
  \quad & = \frac{\mathbf{b}_{\bullet1}^T\mathbf{b}_{\bullet2}}{\sqrt{\mathbf{b}_{\bullet 1}^T\mathbf{b}_{\bullet 1}}\sqrt{\mathbf{b}_{\bullet 2}^T\mathbf{b}_{\bullet 2}}} \nonumber \\
  \quad & = \frac{\langle{\mathbf{b}_{\bullet 1}, \mathbf{b}_{\bullet 2}}\rangle}{\left\lVert \mathbf{b}_{\bullet1}\right\rVert_2 \cdot \left\lVert \mathbf{b}_{\bullet2}\right\rVert_2} \nonumber \\
   \textrm{corr}_{\mathbf{y}_{\bullet1},\mathbf{y}_{\bullet2}} \quad & =  \cos\theta
\end{align}

where \(\mathbf{U}^T \mathbf{U}=\mathbf{I}\) is a property of the
``left'' eigenvectors matrix from Equation \ref{eq:svdbiplot}. In our
example, one can verify that the cosine similarity between
\(\mathbf{b}_{\bullet1}\) and \(\mathbf{b}_{\bullet2}\) is
\(\cos\theta=0.84\), which indeed equivalent to the correlation
coefficient between \(\mathbf{y}_{\bullet1}\) and
\(\mathbf{y}_{\bullet2}\):

\begin{Shaded}
\begin{Highlighting}[]
\NormalTok{  cos\_theta\_ab }\OtherTok{\textless{}{-}}\NormalTok{ (B[}\DecValTok{1}\NormalTok{,] }\SpecialCharTok{\%*\%}\NormalTok{ B[}\DecValTok{2}\NormalTok{,]) }\SpecialCharTok{/}\NormalTok{ (}\FunctionTok{norm}\NormalTok{(B[}\DecValTok{1}\NormalTok{,], }\StringTok{"2"}\NormalTok{)}\SpecialCharTok{*}\FunctionTok{norm}\NormalTok{(B[}\DecValTok{2}\NormalTok{,], }\StringTok{"2"}\NormalTok{))}
  \FunctionTok{all.equal}\NormalTok{(}\FunctionTok{as.numeric}\NormalTok{(cos\_theta\_ab), }\FunctionTok{as.numeric}\NormalTok{(}\FunctionTok{cor}\NormalTok{(Y[,}\DecValTok{1}\NormalTok{], Y[,}\DecValTok{2}\NormalTok{])))}
\end{Highlighting}
\end{Shaded}

\begin{verbatim}
## [1] TRUE
\end{verbatim}

\hypertarget{discussion-of-comparative-insights}{%
\section{Discussion of comparative
insights}\label{discussion-of-comparative-insights}}

This section develops the computational building blocks summarised
earlier, into an implementation-agnostic evaluation grid, and discusses
how specific implementations compare within such grid. The
implementations examined here are those listed in Tab.\ref{tab:tab1pdf},
with key comparative insights summarised upfront in
Tab.\ref{tab:tab3pdf} and Tab.\ref{tab:tab4pdf} below. Within these
tables, the proposed evaluating grid informs the chosen column headers.
The following subsections detail how these insights came about.

\begin{landscape}\begin{table}

\caption{\label{tab:tab3pdf}Selected implementations, PCA focus}
\centering
\begin{tabular}[t]{lllllllllc}
\toprule
\multicolumn{1}{l}{ } & \multicolumn{1}{l}{ } & \multicolumn{1}{l}{ } & \multicolumn{2}{l}{Loadings (eigenvectors)} & \multicolumn{2}{l}{Eigen/singular values} & \multicolumn{3}{l}{PCA scores} \\
\cmidrule(l{3pt}r{3pt}){4-5} \cmidrule(l{3pt}r{3pt}){6-7} \cmidrule(l{3pt}r{3pt}){8-10}
\multicolumn{1}{l}{ } & \multicolumn{1}{l}{ } & \multicolumn{1}{l}{ } & \multicolumn{1}{l}{Cov. mat.} & \multicolumn{1}{l}{SVD right.} & \multicolumn{1}{l}{Cov. mat.} & \multicolumn{1}{l}{SVD} & \multicolumn{1}{l}{Cov. mat.} & \multicolumn{1}{l}{SVD, left} & \multicolumn{1}{l}{Variance} \\
 & Function & M.D.$^1$ & $\mathbf{V}=\left[  \mathbf{a}_{j} \right]_{n \times m}$ & $|\mathbf{v}_j|=|\mathbf{a}_j|$ & $\boldsymbol{\lambda}=\left[\lambda_j \right]_{n \times 1}$ & $\frac{\ell_j}{\sqrt{n-1}}=\sqrt{\lambda_j}$ & $\mathbf{z}_j=\mathbf{Y}\mathbf{a}_{j}$ & $\mathbf{z}_j=\mathbf{u}_j\ell_j$ & $\textrm{Var} \left[ \mathbf{z}_j \right] = \lambda_j$\\
\midrule
\addlinespace[0.3em]
\multicolumn{10}{l}{\textbf{SVD-based}}\\
\hspace{1em}base-$\texttt{R}$ & $\texttt{prcomp()}$ & a &  & $\texttt{rotation}$ &  & $\texttt{sdev}=\sqrt{\lambda_j}$ &  & $\texttt{x}$ & $\bullet$\\
\hspace{1em}$\texttt{pcaMethods}$ & $\texttt{pca()}$ & a &  & $\texttt{loadings}$ &  & $\texttt{sDev}\neq\sqrt{\lambda_j}$ &  & $\texttt{scores}$ & $\bullet$\\
\hspace{1em}$\texttt{PCAtools}$ & $\texttt{pca()}$ & a &  & $\texttt{loadings}$ &  & $\texttt{sdev}\neq\sqrt{\lambda_j}$ &  & $\texttt{rotated}$ & $\bullet$\\
\hspace{1em}$\texttt{ggbiplot}$ & $\texttt{getsvd()}$ &  &  &  &  &  &  &  & \\
\addlinespace[0.3em]
\multicolumn{10}{l}{\textbf{Eigenproblme}}\\
\hspace{1em}base-$\texttt{R}$ & $\texttt{princomp()}$ & b & $\texttt{loadings}$ &  &  & $\texttt{sdev}=\sqrt{\lambda_j}$ & $\texttt{scores}$ &  & \\
\hspace{1em}$\texttt{ade4}$ & $\texttt{dudi.pca()}$ & b & $\texttt{c1}$ &  &  & $\texttt{eig}=\sqrt{\lambda_j}$ & $\texttt{li}$ &  & \\
\hspace{1em}$\texttt{amap}$ & $\texttt{acp()}$ & c & $\texttt{loadings}$ &  &  & $\texttt{eig}=\sqrt{\lambda_j}$ & $\texttt{scores}$ &  & \\
\hspace{1em}$\texttt{psych}$ & $\texttt{principal()}$ & d & $\texttt{loadings}^2$ &  & $\texttt{values}$ &  & $\texttt{scores}^2$ &  & \\
\addlinespace[0.3em]
\multicolumn{10}{l}{\textbf{Gen. SVD-based}}\\
\hspace{1em}$\texttt{FactoMineR}$ & $\texttt{PCA()}$ & a &  & $\texttt{svd\$V}$ &  & $\frac{\ell_j}{\sqrt{n}}=\texttt{svd\$vs}$ &  & $\texttt{ind\$coord}$ & $\circ$\\
\hspace{1em}$\texttt{PCAmixdata}$ & $\texttt{PCAmix()}$ & a &  & $\texttt{svd\$V}$ &  & $\frac{\ell_j}{\sqrt{n}}=\texttt{svd\$vs}$ &  & $\texttt{ind\$coord}$ & $\circ$\\
\hspace{1em}$\texttt{factoextra}$ & $\texttt{getpca()}$ &  &  &  &  &  &  &  & \\
\bottomrule
\multicolumn{10}{l}{\rule{0pt}{1em}\textit{Note: } The evaluation grid appears in the column headings. $\circ$ indicates an equivalence that holds with caveats.}\\
\multicolumn{10}{l}{\rule{0pt}{1em}\textsuperscript{1} Matrix being decomposed: (a) $\mathbf{Y}$; (b) $n^{-1}\mathbf{Y}^T\mathbf{Y}$; (c) $\mathbf{Y}^T\mathbf{Y}$; (d) $\mathbf{S}$ \textsuperscript{2} These outputs correspond to, respectively: $\mathbf{a}_j\sqrt{\lambda_j}$ and $\mathbf{z}_j\sqrt{\lambda_j}$}\\
\end{tabular}
\end{table}
\end{landscape}

\begin{landscape}\begin{table}

\caption{\label{tab:tab4pdf}Selected implementations, principal component biplots ($\alpha=0$)}
\centering
\begin{tabular}[t]{llcccccc}
\toprule
\multicolumn{1}{l}{ } & \multicolumn{1}{l}{ } & \multicolumn{2}{l}{Observ. coord.} & \multicolumn{2}{l}{Feat. coord.} & \multicolumn{2}{l}{Feat. coord. properties} \\
\cmidrule(l{3pt}r{3pt}){3-4} \cmidrule(l{3pt}r{3pt}){5-6} \cmidrule(l{3pt}r{3pt}){7-8}
\multicolumn{1}{l}{ } & \multicolumn{1}{l}{ } & \multicolumn{1}{l}{Scaled} & \multicolumn{1}{l}{As is$^1$} & \multicolumn{1}{l}{Scaled} & \multicolumn{1}{l}{As is$^1$} & \multicolumn{1}{l}{St. Dev. (single feat.)} & \multicolumn{1}{l}{Corr. coeff. (feat. pair)} \\
  & Function & $\mathbf{A}=\mathbf{U}$ & $\mathbf{A}=\mathbf{Z}$ & $\mathbf{B}^T=\mathbf{D}\mathbf{V}^T$ & $\mathbf{B}=\mathbf{V}$ & $\left\lVert \mathbf{b}_j \right\rVert  (n-1)^{1/2}=\sigma_j$ & $\textrm{corr}_{\mathbf{y}_{i},\mathbf{y}_{j}} =  \textrm{cos}\theta_{ij}$\\
\midrule
\addlinespace[0.3em]
\multicolumn{8}{l}{\textbf{SVD-based}}\\
\hspace{1em}base-$\texttt{R}$ & $\texttt{biplot()}$ & $\circ$ &  & $\circ$ &  & $\circ$ & \vphantom{1} $\circ$\\
\hspace{1em}$\texttt{pcaMethods}$ & $\texttt{slplot()}$ &  & $\bullet$ &  & $\bullet$ &  & \\
\hspace{1em}$\texttt{PCAtools}$ & $\texttt{biplot()}$ &  & $\bullet$ &  & $\bullet$ &  & \\
\hspace{1em}$\texttt{ggbiplot}$ & $\texttt{ggbiplot()}$ &  & $\circ$ & $\circ$ &  & $\circ$ & $\circ$\\
\addlinespace[0.3em]
\multicolumn{8}{l}{\textbf{Eigenproblem}}\\
\hspace{1em}base-$\texttt{R}$ & $\texttt{biplot()}$ & $\circ$ &  & $\circ$ &  & $\circ$ & $\circ$\\
\hspace{1em}$\texttt{ade4}$ & $\texttt{scatter()}$ &  & $\bullet$ &  & $\bullet$ &  & \\
\hspace{1em}$\texttt{amap}$ & $\texttt{plot()}$ &  & $\bullet$ &  & $\bullet$ &  & \\
\hspace{1em}$\texttt{psych}$ & $\texttt{biplot.psych()}$ &  & $\bullet$ &  & $\bullet$ &  & \\
\addlinespace[0.3em]
\multicolumn{8}{l}{\textbf{Gen. SVD-based}}\\
\hspace{1em}$\texttt{FactoMineR}$ & $\texttt{plot.PCA()}$ &  & $\bullet$ & $\bullet$ &  &  & \\
\hspace{1em}$\texttt{PCAmixdata}$ & $\texttt{plot.PCAmix()}$ &  &  & $\bullet$ &  &  & \\
\hspace{1em}$\texttt{factoextra}$ & $\texttt{fviz\_pca\_biplot()}$ &  & $\bullet$ & $\bullet$ &  &  & \\
\bottomrule
\multicolumn{8}{l}{\rule{0pt}{1em}\textit{Note: } The evaluation grid appears in the column headings. $\circ$ indicates an equivalence that holds with caveats.}\\
\multicolumn{8}{l}{\rule{0pt}{1em}\textsuperscript{1} Equivalent to the PCA scores matrix; \textsuperscript{2} Equivalent to the loadings matrix.}\\
\end{tabular}
\end{table}
\end{landscape}

\begin{figure}

{\centering \includegraphics[width=0.5\linewidth]{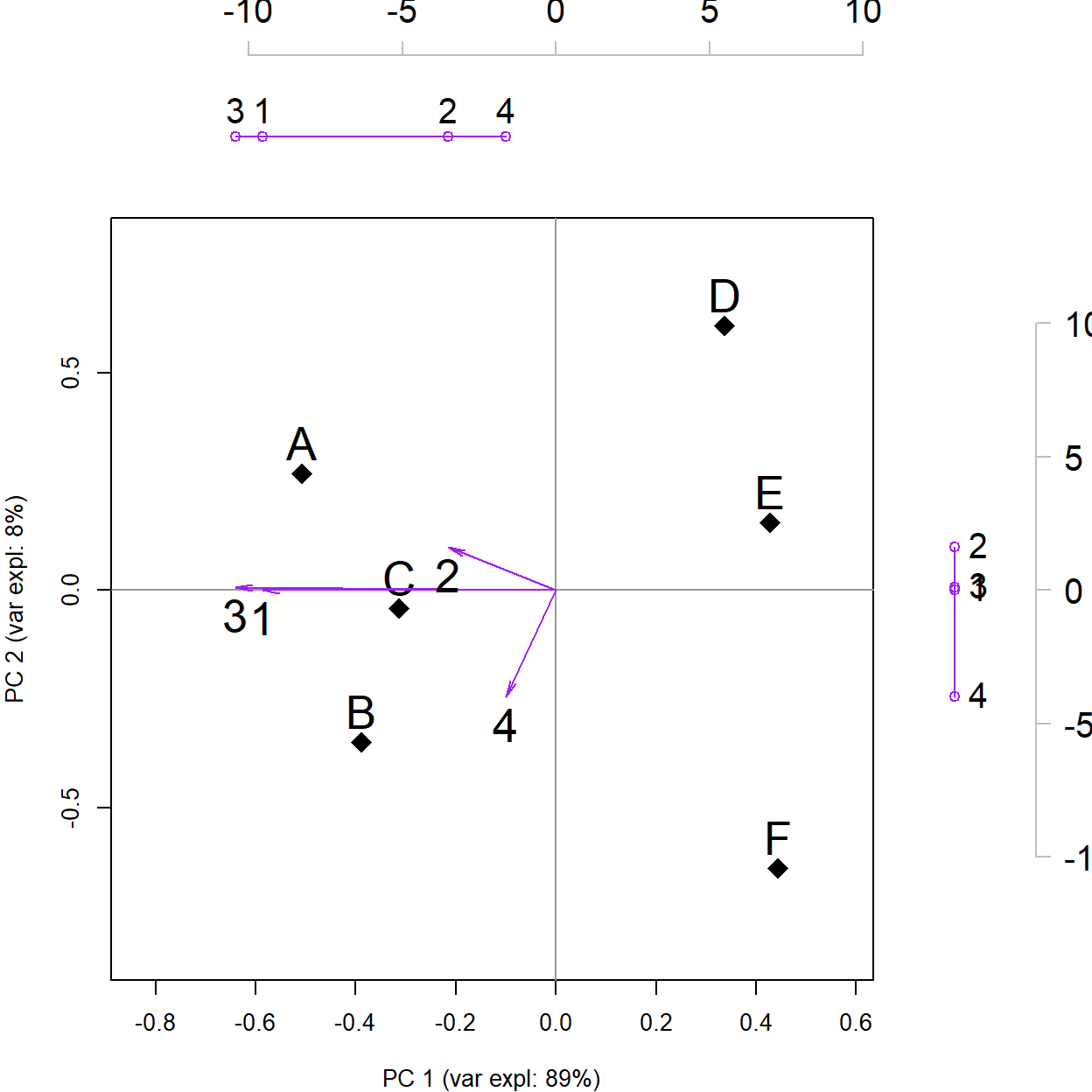} 

}

\caption{Biplot of observations and features (shown as numbers) for an extended example based on SVD. The superposed dual axes outside the plot reflect the different scale of the features' coordinates.}\label{fig:fig3pdf}
\end{figure}

To provide a benchmark, Tab.\ref{tab:tab5pdf} extends the illustrative
example to include more than two features. Fig.\ref{fig:fig3pdf} shows
the corresponding biplot for a rank-2 approximation assuming that the
first two components are retained. The tables show that specific
functions may share the same algebraic rationale, yet differ in how the
relevant computations are carried out for each building block.

The grid includes verification equivalences corresponding to Equations
\ref{eq:eigvar}, \ref{eq:length}, and \ref{eq:coscorr} i.e.,
relationships that should hold if the computational rationale
underpinning each technique---illustrated in previous sections---is
followed correctly. Strikingly, these equivalences rarely follow without
caveats from the output of specific implementations alone.

\begin{table}
\centering
\caption{\label{tab:tab5pdf}PCA scores for an illustrative example with $m=4$ features and $n=6$ observations. Matrix $\mathbf{Z}$ obtained by eigendecomposition of the covariance matrix $\mathbf{S}$; matrix $\mathbf{Z}_{SVD}$ obtained by SVD of matrix $\mathbf{Y}$; Last two columns: observations' coordinates on the biplot also from SVD}
\fontsize{7}{9}\selectfont
\resizebox{\ifdim\width>\linewidth\linewidth\else\width\fi}{!}{
\begin{tabular}[t]{lrrrrrrrrrrrrrr}
\toprule
\multicolumn{1}{l}{ } & \multicolumn{4}{l}{Raw data $\mathbf{X}$} & \multicolumn{4}{l}{PCA scores $\mathbf{Z}_{\textrm{SVD}}=\mathbf{UD}$} & \multicolumn{4}{l}{PCA scores $\mathbf{Z}$} & \multicolumn{2}{l}{Biplot $\mathbf{U}$} \\
\cmidrule(l{3pt}r{3pt}){2-5} \cmidrule(l{3pt}r{3pt}){6-9} \cmidrule(l{3pt}r{3pt}){10-13} \cmidrule(l{3pt}r{3pt}){14-15}
  & feat1 & feat2 & feat3 & feat4 & PC1 & PC2 & PC3 & PC4 & PC1 & PC2 & PC3 & PC4 & PC1 & PC2\\
\midrule
A & 10.00 & 6.00 & 12.00 & 5.00 & -7.44 & 1.16 & -0.89 & -0.04 & 7.44 & -1.16 & -0.89 & -0.04 & -0.51 & 0.27\\
B & 11.00 & 4.00 & 9.00 & 7.00 & -5.70 & -1.51 & 1.92 & -0.03 & 5.70 & 1.51 & 1.92 & -0.03 & -0.39 & -0.35\\
C & 8.00 & 5.00 & 10.00 & 6.00 & -4.58 & -0.18 & -1.08 & 0.13 & 4.58 & 0.18 & -1.08 & 0.13 & -0.31 & -0.04\\
D & 3.00 & 3.00 & 2.50 & 2.00 & 4.94 & 2.63 & 0.51 & -0.38 & -4.94 & -2.63 & 0.51 & -0.38 & 0.34 & 0.61\\
E & 2.00 & 2.80 & 1.30 & 4.00 & 6.26 & 0.67 & 0.32 & 0.55 & -6.26 & -0.67 & 0.32 & 0.55 & 0.43 & 0.16\\
\addlinespace
F & 1.00 & 1.00 & 2.00 & 7.00 & 6.51 & -2.77 & -0.79 & -0.23 & -6.51 & 2.77 & -0.79 & -0.23 & 0.44 & -0.64\\
\midrule
mean & 5.83 & 3.63 & 6.13 & 5.17 & 0.00 & 0.00 & 0.00 & 0.00 & 0.00 & 0.00 & 0.00 & 0.00 & 0.00 & 0.00\\
sample var. & 18.97 & 3.13 & 22.25 & 3.77 & 42.95 & 3.73 & 1.32 & 0.10 & 42.95 & 3.73 & 1.32 & 0.10 & 0.20 & 0.20\\
\bottomrule
\end{tabular}}
\end{table}

\hypertarget{group-1-svd}{%
\subsection{Group 1: SVD}\label{group-1-svd}}

The first group of \texttt{R} implementations considered here has a
common trait: the function \texttt{svd()} is at work under the hood,
providing what is needed to obtain a \emph{scores matrix}, as per
Equation \ref{eq:svdscores}, as well as the coordinates for both
observations and features in a principal component biplot, as per
Equation \ref{eq:svdbiplot}. This is specifically the case for
Base-\texttt{R}, functions like \texttt{prcomp()} and \texttt{biplot()},
although some caveats apply in the latter case.

For example, the command
\texttt{test\_pca1\textless{}-prcomp(X,\ scale\ =\ FALSE)} applies
\texttt{svd()} under the hood to the centred data matrix \(\mathbf{Y}\).
The returned output \texttt{x} corresponds to the PCA \emph{scores} in
columns 7-8 in Tab.\ref{tab:tab5pdf}, whereas \texttt{rotation}, returns
the matrix of right eigenvectors or \emph{loadings}.

One aspect that stands out from Tab.\ref{tab:tab3pdf} is that, unlike
other implementations in this group, \texttt{prcomp()} satisfies
Equation \ref{eq:svev}, which verifies the relationship between the
eigenvalues of the covariance matrix \(\mathbf{S}\), the St.Dev. of the
scores, and the singular values from an SVD:

\begin{Shaded}
\begin{Highlighting}[]
\FunctionTok{all.equal}\NormalTok{(test\_pca1}\SpecialCharTok{$}\NormalTok{sdev}\SpecialCharTok{*}\FunctionTok{sqrt}\NormalTok{(}\FunctionTok{nrow}\NormalTok{(Y)}\SpecialCharTok{{-}}\DecValTok{1}\NormalTok{), }\FunctionTok{svd}\NormalTok{(Y)}\SpecialCharTok{$}\NormalTok{d)}
\end{Highlighting}
\end{Shaded}

\begin{verbatim}
## [1] TRUE
\end{verbatim}

As a corollary, the function also satisfies Equation \ref{eq:eigvar}:

\begin{Shaded}
\begin{Highlighting}[]
\FunctionTok{all.equal}\NormalTok{(}\FunctionTok{as.numeric}\NormalTok{(test\_pca1}\SpecialCharTok{$}\NormalTok{sdev}\SpecialCharTok{\^{}}\DecValTok{2}\NormalTok{), }\FunctionTok{as.numeric}\NormalTok{(}\FunctionTok{apply}\NormalTok{(test\_pca1}\SpecialCharTok{$}\NormalTok{x,}\DecValTok{2}\NormalTok{,var)))}
\end{Highlighting}
\end{Shaded}

\begin{verbatim}
## [1] TRUE
\end{verbatim}

The script \texttt{stats:::prcomp.default} reveals how this is achieved:
the output \texttt{sdev} returns the singular values from the SVD
divided by \(\sqrt{n-1}\), consistently with Equation \ref{eq:svev}:

\begin{Shaded}
\begin{Highlighting}[]
\FunctionTok{all.equal}\NormalTok{(}\FunctionTok{svd}\NormalTok{(Y)}\SpecialCharTok{$}\NormalTok{d}\SpecialCharTok{/}\FunctionTok{sqrt}\NormalTok{(}\FunctionTok{nrow}\NormalTok{(Y)}\SpecialCharTok{{-}}\DecValTok{1}\NormalTok{), test\_pca1}\SpecialCharTok{$}\NormalTok{sdev)}
\end{Highlighting}
\end{Shaded}

\begin{verbatim}
## [1] TRUE
\end{verbatim}

In base-\texttt{R}, a principal component biplot can be obtained from
\texttt{prcomp()} via the \texttt{biplot()} function. Although users are
presented with a graphical output, \texttt{biplot()} does not return the
jointly computed coordinates for features and observations. The
function's script \texttt{stats:::biplot.prcomp} reveals good alignment
with Equations \ref{eq:obscoordbiplot} and \ref{eq:pcasd}, except for an
imprecision in how the singular values are arrived at---which explains
the caveats for this function in Tab.\ref{tab:tab3pdf}.

Specifically, \texttt{biplot.prcomp} computes the singular values from
the output \texttt{sdev} of a \texttt{prcomp} object. How it does so
depends on the argument \texttt{pc.biplot}: the default setting is
\texttt{FALSE}, with \texttt{TRUE} denoting that a principal component
biplot is desired. Contrary to what one may expect, when
\texttt{pc.biplot=TRUE} the function sets the singular values equal to
\texttt{sdev}, disagreeing with Equation \ref{eq:svev}, which in turn
affects Equations \ref{eq:obscoordbiplot}, \ref{eq:pcasd} and
\ref{eq:length}. Surprisingly, when \texttt{pc.biplot=FALSE} the
singular values are computed in a fashion similar to Equation
\ref{eq:svev}, although it is erroneously assumed that \texttt{sdev}
returns the singular values divided by \(\sqrt{n}\) instead of
\(\sqrt{n-1}\), the latter being the case for \texttt{prcomp()}.

Due to these assumptions, when the argument \texttt{pc.biplot} is
(contestably) \texttt{FALSE} the singular values returned by
\texttt{biplot.prcomp} are close to, but different from
\(\boldsymbol{\ell}\) from Equation \ref{eq:svev}:

\begin{Shaded}
\begin{Highlighting}[]
\NormalTok{lam }\OtherTok{\textless{}{-}}\NormalTok{ test\_pca1}\SpecialCharTok{$}\NormalTok{sdev}\SpecialCharTok{*}\FunctionTok{sqrt}\NormalTok{(}\FunctionTok{nrow}\NormalTok{(test\_pca1}\SpecialCharTok{$}\NormalTok{x))}
\FunctionTok{all.equal}\NormalTok{(lam, }\FunctionTok{svd}\NormalTok{(Y)}\SpecialCharTok{$}\NormalTok{d)}
\end{Highlighting}
\end{Shaded}

\begin{verbatim}
## [1] "Mean relative difference: 0.08712907"
\end{verbatim}

With the singular values thus obtained, \texttt{biplot.prcomp} computes
the matrix of observations coordinates \(\mathbf{A}\) from the scores
matrix consistently with Equation \ref{eq:obscoordbiplot}. Yet the
result differs from what Equation \ref{eq:obscoordbiplot} suggests, and
hence from what is shown in Tab.\ref{tab:tab5pdf}and
Fig.\ref{fig:fig3pdf}:

\begin{Shaded}
\begin{Highlighting}[]
\FunctionTok{all.equal}\NormalTok{(}\FunctionTok{unname}\NormalTok{(}\FunctionTok{sweep}\NormalTok{(test\_pca1}\SpecialCharTok{$}\NormalTok{x,}\DecValTok{2}\NormalTok{,lam,}\StringTok{"/"}\NormalTok{)), }\FunctionTok{svd}\NormalTok{(Y)}\SpecialCharTok{$}\NormalTok{u)}
\end{Highlighting}
\end{Shaded}

\begin{verbatim}
## [1] "Mean relative difference: 0.09544512"
\end{verbatim}

The caveat in Tab.\ref{tab:tab4pdf} acknowledges that the equivalence
\(\mathbf{A} = \mathbf{U}\) could be attained with minor changes in how
singular values are computed by \texttt{biplot.prcomp}, namely:

\begin{Shaded}
\begin{Highlighting}[]
\NormalTok{sv }\OtherTok{\textless{}{-}}\NormalTok{ test\_pca1}\SpecialCharTok{$}\NormalTok{sdev}\SpecialCharTok{*}\FunctionTok{sqrt}\NormalTok{(}\FunctionTok{nrow}\NormalTok{(Y)}\SpecialCharTok{{-}}\DecValTok{1}\NormalTok{)}
\FunctionTok{all.equal}\NormalTok{(}\FunctionTok{unname}\NormalTok{(}\FunctionTok{sweep}\NormalTok{(test\_pca1}\SpecialCharTok{$}\NormalTok{x,}\DecValTok{2}\NormalTok{,sv,}\StringTok{"/"}\NormalTok{)), }\FunctionTok{unname}\NormalTok{(}\FunctionTok{svd}\NormalTok{(Y)}\SpecialCharTok{$}\NormalTok{u))}
\end{Highlighting}
\end{Shaded}

\begin{verbatim}
## [1] TRUE
\end{verbatim}

Similar caveats apply to the features' coordinates. In principle, the
underlying computations in \texttt{biplot.prcomp} are consistent with
Equation \ref{eq:pcasd}, which yields matrix \(\mathbf{B}\), except that
these results are also affected by how singular values are computed. In
our example, the features' coordinates produced by \texttt{biplot()} are
different from matrix \(\mathbf{B}\) in Equation \ref{eq:pcasd}---which
is displayed in Fig.\ref{fig:fig3pdf}:

\begin{Shaded}
\begin{Highlighting}[]
\NormalTok{B }\OtherTok{\textless{}{-}} \FunctionTok{t}\NormalTok{(D }\SpecialCharTok{\%*\%} \FunctionTok{t}\NormalTok{(V))}
\FunctionTok{all.equal}\NormalTok{(}\FunctionTok{unname}\NormalTok{(}\FunctionTok{sweep}\NormalTok{(test\_pca1}\SpecialCharTok{$}\NormalTok{rotation,}\DecValTok{2}\NormalTok{,lam,}\StringTok{"*"}\NormalTok{)), B)}
\end{Highlighting}
\end{Shaded}

\begin{verbatim}
## [1] "Mean relative difference: 0.08712907"
\end{verbatim}

Just like the base-\texttt{R} functions examined so far, several curated
packages rely on conventional SVD, too. In particular,
\emph{\href{https://bioconductor.org/packages/3.18/PcaMethods}{PcaMethods}}
(\protect\hyperlink{ref-pcaMethods_paper}{Stacklies et al. 2007}) and
\emph{\href{https://bioconductor.org/packages/3.18/PCAtools}{PCAtools}}
(\protect\hyperlink{ref-PCAtools}{Blighe and Lun 2023}) extend
traditional PCA methods to handle computational challenges that are
specific to bioinformatics data. While both packages offer a homonym
function \texttt{pca()}, they differ in the extent to which they rely on
base-\texttt{R} function. For the implementation
\texttt{pcaMethods::svdPca}, reliance on \texttt{prcomp()} is explicit
upon examination of the function's script, and promptly ascertained. For
example:

\begin{Shaded}
\begin{Highlighting}[]
\NormalTok{test\_pca2 }\OtherTok{\textless{}{-}}\NormalTok{ pcaMethods}\SpecialCharTok{::}\FunctionTok{pca}\NormalTok{(Y, }\AttributeTok{method =} \StringTok{"svd"}\NormalTok{)}
\FunctionTok{all.equal}\NormalTok{(test\_pca2}\SpecialCharTok{@}\NormalTok{scores, test\_pca1}\SpecialCharTok{$}\NormalTok{x[,}\DecValTok{1}\SpecialCharTok{:}\DecValTok{2}\NormalTok{] )}
\end{Highlighting}
\end{Shaded}

\begin{verbatim}
## [1] TRUE
\end{verbatim}

So long as the selected method in \texttt{pcaMethods::svdPca} is
\texttt{svd} the equivalence with \texttt{prcomp()} can be verified in a
similar fashion for \texttt{loadings} and \texttt{sDev}.

With regards to \texttt{PCAtools::pca}, the analogy with
\texttt{prcomp()} is less straightforward to identify as it relies on
another package,
\emph{\href{https://bioconductor.org/packages/3.18/BiocSingular}{BiocSingular}}
, to carry out an SVD. Yet the results associated with a so-called
``exact'' SVD are, once again, equivalent to \texttt{prcomp()}. For
example:

\begin{Shaded}
\begin{Highlighting}[]
\NormalTok{test\_pca3 }\OtherTok{\textless{}{-}}\NormalTok{ PCAtools}\SpecialCharTok{::}\FunctionTok{pca}\NormalTok{(X, }\AttributeTok{transposed =}\NormalTok{ T)}
\FunctionTok{all.equal}\NormalTok{(}\FunctionTok{as.matrix}\NormalTok{(test\_pca3}\SpecialCharTok{$}\NormalTok{rotated), test\_pca1}\SpecialCharTok{$}\NormalTok{x)}
\end{Highlighting}
\end{Shaded}

\begin{verbatim}
## [1] TRUE
\end{verbatim}

The equivalence with \texttt{prcomp()} can be verified in a similar
fashion for \texttt{loadings} and \texttt{sdev}.

Unlike their base-\texttt{R} counterpart, the biplot functions in
packages
\emph{\href{https://bioconductor.org/packages/3.18/pcaMethods}{pcaMethods}}
and
\emph{\href{https://bioconductor.org/packages/3.18/PCAtools}{PCAtools}}
do not strictly speaking generate a principal components biplot as per
Equation \ref{eq:svdbiplot}. The function \texttt{pcaMethods::slplot}
displays two separate plots of \emph{scores} and \emph{loadings} whose
coordinates are derived, respectively, from he the outputs \texttt{x}
and \texttt{rotation} of the function \texttt{prcomp()} without further
processing. Despite its name, the function \texttt{PCAtools::biplot()},
too, overlays a scores and a loadings plot.

The packages
\emph{\href{https://CRAN.R-project.org/package=ggbiplot}{ggbiplot}}
(\protect\hyperlink{ref-ggbiplot_vignette}{Vu and Friendly 2024}) is
dedicated to biplots. Like \texttt{biplot()} in base-\texttt{R}, also
\texttt{ggbiplot()} assumes that an SVD-based PCA be carried out
separately e.g., by \texttt{prcomp()}. The documentation acknowledges
that PCA implementations may differ in how they go about an SVD, and
that \texttt{get\_SVD()} seeks to provide a unifying interface. Yet
\texttt{get\_SVD()} suffers from some imprecisions already encountered
for \texttt{biplot.prcomp} e.g., it assumes that the output
\texttt{sdev} of a \texttt{prcomp()} object is equivalent to the
singular values divided by \(\sqrt{n}\) instead of \(\sqrt{n-1}\). For
the singular values thus defined, both \texttt{get\_SVD()} and
\texttt{biplot.prcomp} return the same observations' coordinates:

\begin{Shaded}
\begin{Highlighting}[]
\NormalTok{test\_pca4 }\OtherTok{\textless{}{-}}\NormalTok{ ggbiplot}\SpecialCharTok{::}\FunctionTok{get\_SVD}\NormalTok{(test\_pca1)}
\FunctionTok{all.equal}\NormalTok{(test\_pca4}\SpecialCharTok{$}\NormalTok{U[,}\DecValTok{1}\SpecialCharTok{:}\DecValTok{2}\NormalTok{], }\FunctionTok{sweep}\NormalTok{(test\_pca1}\SpecialCharTok{$}\NormalTok{x,}\DecValTok{2}\NormalTok{,lam,}\StringTok{"/"}\NormalTok{)[,}\DecValTok{1}\SpecialCharTok{:}\DecValTok{2}\NormalTok{])}
\end{Highlighting}
\end{Shaded}

\begin{verbatim}
## [1] TRUE
\end{verbatim}

If Equation \ref{eq:svdbiplot} was followed, instead, the output
\texttt{U} from \texttt{ggbiplot::get\_SVD()} would represent the biplot
coordinates for the observations, as well as the matrix of left
eigenvalues from an SVD. Instead, such output is further processed by
the function \texttt{ggbiplot()} as follows: if the desired output is a
principal component biplot i.e., if the argument \texttt{pc.biplot} is
true, \texttt{ggbiplot()} multiplies \texttt{U} by \texttt{sdev} and,
continuing with assumpion the input is a \texttt{prcomp()} object, by
\(\sqrt{n-1}\) as well. Ultimately, this process reverts back to the PCA
scores, multiplied by the ratio \(\sqrt{n-1}/ \sqrt{n}\), hence the
caveat in Tab.\ref{tab:tab4pdf}:

\begin{Shaded}
\begin{Highlighting}[]
\NormalTok{df.u }\OtherTok{\textless{}{-}} \FunctionTok{as.data.frame}\NormalTok{(}\FunctionTok{sweep}\NormalTok{(test\_pca4}\SpecialCharTok{$}\NormalTok{U[, }\DecValTok{1}\SpecialCharTok{:}\DecValTok{2}\NormalTok{], }\DecValTok{2}\NormalTok{, test\_pca4}\SpecialCharTok{$}\NormalTok{D[}\DecValTok{1}\SpecialCharTok{:}\DecValTok{2}\NormalTok{], }\AttributeTok{FUN =} \StringTok{"*"}\NormalTok{))}
\NormalTok{df.u }\OtherTok{\textless{}{-}}\NormalTok{ df.u }\SpecialCharTok{*} \FunctionTok{sqrt}\NormalTok{(}\FunctionTok{nrow}\NormalTok{(test\_pca1}\SpecialCharTok{$}\NormalTok{x) }\SpecialCharTok{{-}}  \DecValTok{1}\NormalTok{)}
\FunctionTok{all.equal}\NormalTok{(test\_pca1}\SpecialCharTok{$}\NormalTok{x[,}\DecValTok{1}\SpecialCharTok{:}\DecValTok{2}\NormalTok{] }\SpecialCharTok{*} \FunctionTok{sqrt}\NormalTok{(}\FunctionTok{nrow}\NormalTok{(Y)}\SpecialCharTok{{-}}\DecValTok{1}\NormalTok{) }\SpecialCharTok{/} \FunctionTok{sqrt}\NormalTok{(}\FunctionTok{nrow}\NormalTok{(Y)), }\FunctionTok{as.matrix}\NormalTok{(df.u))}
\end{Highlighting}
\end{Shaded}

\begin{verbatim}
## [1] TRUE
\end{verbatim}

Moving on to the features' coordinates, \texttt{ggbiplot()} operates
similarly to Equation \ref{eq:pcasd}, except that the scores' St.Dev.
\texttt{sdev} from \texttt{prcomp()} are used instead of the singular
values \(\mathbf{D}\). Hence the results not only disagree with Equation
\ref{eq:pcasd}, but also with \texttt{biplot.prcomp}:

\begin{Shaded}
\begin{Highlighting}[]
\NormalTok{df.v }\OtherTok{\textless{}{-}} \FunctionTok{as.data.frame}\NormalTok{(}\FunctionTok{sweep}\NormalTok{(test\_pca1}\SpecialCharTok{$}\NormalTok{rotation[, }\DecValTok{1}\SpecialCharTok{:}\DecValTok{2}\NormalTok{], }\DecValTok{2}\NormalTok{, test\_pca4}\SpecialCharTok{$}\NormalTok{D[}\DecValTok{1}\SpecialCharTok{:}\DecValTok{2}\NormalTok{], }\AttributeTok{FUN =} \StringTok{"*"}\NormalTok{))}
\FunctionTok{all.equal}\NormalTok{(}\FunctionTok{sweep}\NormalTok{(test\_pca1}\SpecialCharTok{$}\NormalTok{rotation[,}\DecValTok{1}\SpecialCharTok{:}\DecValTok{2}\NormalTok{],}\DecValTok{2}\NormalTok{,lam[}\DecValTok{1}\SpecialCharTok{:}\DecValTok{2}\NormalTok{],}\StringTok{"*"}\NormalTok{), }\FunctionTok{as.matrix}\NormalTok{(df.v))}
\end{Highlighting}
\end{Shaded}

\begin{verbatim}
## [1] "Mean relative difference: 0.5917517"
\end{verbatim}

\hypertarget{group-2-eigenvalue-problem}{%
\subsection{Group 2: eigenvalue
problem}\label{group-2-eigenvalue-problem}}

The second group of implementations considered here follows more closely
the variance optimisation rationale underlying PCA, and the
eigendecomposition problem in Equation \ref{eq:eigprob} concerning the
covariance matrix \(\mathbf{S}\). Whilst less prevalent than SVD,
\texttt{R} implementations adopting this approach are commonly adopted
in the literature---e.g., Mayor
(\protect\hyperlink{ref-MayorEric2015Lpaw}{2015}); Hanson and Harvey
(\protect\hyperlink{ref-LearnPCA}{2022}). Yet there may be discrepancies
between our agnostic evaluation grid for both base-\texttt{R} and
curated packages.

A natural starting point is the function \texttt{princomp()} in
base-\texttt{R}. Nearly a homonym of \texttt{prcomp()}, this
implementation relies upon the eigenvalues and eigenvectors of the
covariance matrix \(\mathbf{S}\), rather than the outputs of a
data-matrix SVD. Yet \texttt{princomp()} disagrees with Equation
\ref{eq:var} in that it uses the number of observations \(n\), not
\(n-1\), as the divisor for the covariance matrix. What is more, the
signs of the scores returned by \texttt{princomp()} differ from those
computed directly using \texttt{eig()} due to further processing within
the function:

\begin{Shaded}
\begin{Highlighting}[]
\NormalTok{test\_pca5 }\OtherTok{\textless{}{-}} \FunctionTok{princomp}\NormalTok{(X, }\AttributeTok{fix\_sign =}\NormalTok{ F)}
\NormalTok{test\_pca5\_eig }\OtherTok{\textless{}{-}} \FunctionTok{eigen}\NormalTok{(}\DecValTok{1}\SpecialCharTok{/}\FunctionTok{nrow}\NormalTok{(Y)}\SpecialCharTok{*}\FunctionTok{t}\NormalTok{(Y) }\SpecialCharTok{\%*\%}\NormalTok{ Y, }\AttributeTok{symmetric =} \ConstantTok{TRUE}\NormalTok{)}\SpecialCharTok{$}\NormalTok{vectors}
\FunctionTok{all.equal}\NormalTok{(}\FunctionTok{abs}\NormalTok{(}\FunctionTok{unname}\NormalTok{(Y }\SpecialCharTok{\%*\%}\NormalTok{ test\_pca5\_eig)), }\FunctionTok{abs}\NormalTok{(}\FunctionTok{unname}\NormalTok{(test\_pca5}\SpecialCharTok{$}\NormalTok{scores)))}
\end{Highlighting}
\end{Shaded}

\begin{verbatim}
## [1] TRUE
\end{verbatim}

Hence, the scores returned by \texttt{princomp()} differ in sign from
\(\mathbf{Z}\) shown in Tab.\ref{tab:tab5pdf} despite both being
obtained from \texttt{eig()}.

When PCA is implemented using \texttt{princomp()} instead of
\texttt{prcomp()}, the eigenvalues do not equal the variance of the
scores as in Equation \ref{eq:eigvar}:

\begin{Shaded}
\begin{Highlighting}[]
 \FunctionTok{all.equal}\NormalTok{(}\FunctionTok{as.numeric}\NormalTok{(test\_pca5}\SpecialCharTok{$}\NormalTok{sdev}\SpecialCharTok{\^{}}\DecValTok{2}\NormalTok{), }\FunctionTok{as.numeric}\NormalTok{(}\FunctionTok{apply}\NormalTok{(test\_pca5}\SpecialCharTok{$}\NormalTok{scores,}\DecValTok{2}\NormalTok{,var)))}
\end{Highlighting}
\end{Shaded}

\begin{verbatim}
## [1] "Mean relative difference: 0.2"
\end{verbatim}

An object form \texttt{princomp()} can serve as an input to the
functions \texttt{biplot()} and \texttt{ggbiplot::ggbiplot()} examined
earlier. Hence, the same caveats discussed before apply, with the
difference that now both functions are correct in assuming that the the
singular values from an SVD are equivalent to multiplying the output
\texttt{sdev} of \texttt{princomp()} times \(\sqrt{n}\):

\begin{Shaded}
\begin{Highlighting}[]
\FunctionTok{all.equal}\NormalTok{(}\FunctionTok{as.numeric}\NormalTok{(}\FunctionTok{svd}\NormalTok{(Y)}\SpecialCharTok{$}\NormalTok{d), }\FunctionTok{as.numeric}\NormalTok{(test\_pca5}\SpecialCharTok{$}\NormalTok{sdev}\SpecialCharTok{*}\FunctionTok{sqrt}\NormalTok{(}\FunctionTok{nrow}\NormalTok{(Y))))}
\end{Highlighting}
\end{Shaded}

\begin{verbatim}
## [1] TRUE
\end{verbatim}

An option which is very close to \texttt{princomp()} in base-\texttt{R}
is the function \texttt{dudi.pca()} from the package
\emph{\href{https://CRAN.R-project.org/package=ade4}{ade4}}(\protect\hyperlink{ref-ade4book}{Thioulouse
et al. 2018}). Both functions define the covariance matrix of the
centred data-points in similar terms, thus generating comparable outputs
for PCA scores

\begin{Shaded}
\begin{Highlighting}[]
\NormalTok{test\_pca6 }\OtherTok{\textless{}{-}}\NormalTok{ ade4}\SpecialCharTok{::}\FunctionTok{dudi.pca}\NormalTok{(X, }\AttributeTok{center =} \ConstantTok{TRUE}\NormalTok{, }\AttributeTok{scale =} \ConstantTok{FALSE}\NormalTok{, }\AttributeTok{scannf =} \ConstantTok{FALSE}\NormalTok{, }\AttributeTok{nf =} \FunctionTok{ncol}\NormalTok{(Y))}
\FunctionTok{all.equal}\NormalTok{(}\FunctionTok{unname}\NormalTok{(}\FunctionTok{as.matrix}\NormalTok{(test\_pca6}\SpecialCharTok{$}\NormalTok{li)), }\FunctionTok{unname}\NormalTok{(}\FunctionTok{as.matrix}\NormalTok{(test\_pca5}\SpecialCharTok{$}\NormalTok{scores)))}
\end{Highlighting}
\end{Shaded}

\begin{verbatim}
## [1] TRUE
\end{verbatim}

Similar equivalences hold for the eigenvectors (loadings) and
eigenvalues of covariance matrix. The implementation in
\texttt{dudi.pca()}, too, disagrees with Equation \ref{eq:eigvar}:

\begin{Shaded}
\begin{Highlighting}[]
 \FunctionTok{all.equal}\NormalTok{(}\FunctionTok{as.numeric}\NormalTok{(test\_pca6}\SpecialCharTok{$}\NormalTok{eig), }\FunctionTok{as.numeric}\NormalTok{(}\FunctionTok{apply}\NormalTok{(test\_pca6}\SpecialCharTok{$}\NormalTok{li,}\DecValTok{2}\NormalTok{,var)))}
\end{Highlighting}
\end{Shaded}

\begin{verbatim}
## [1] "Mean relative difference: 0.2"
\end{verbatim}

Package \emph{\href{https://CRAN.R-project.org/package=amap}{amap}}
(\protect\hyperlink{ref-amap}{Lucas 2022}) provides a different take on
the covariance matrix with its function \texttt{amap:::acp()}, which
computes the eigenvalues and eigenvectors of \(\mathbf{Y}^T\mathbf{Y}\),
unlike the previous options:

\begin{Shaded}
\begin{Highlighting}[]
\NormalTok{test\_pca7 }\OtherTok{\textless{}{-}}\NormalTok{ amap}\SpecialCharTok{:::}\FunctionTok{acp}\NormalTok{(X, }\AttributeTok{reduce =} \ConstantTok{FALSE}\NormalTok{)}
\NormalTok{test\_pca7\_eig }\OtherTok{\textless{}{-}} \FunctionTok{eigen}\NormalTok{(}\FunctionTok{t}\NormalTok{(Y) }\SpecialCharTok{\%*\%}\NormalTok{ Y, }\AttributeTok{symmetric =} \ConstantTok{FALSE}\NormalTok{)}
\FunctionTok{all.equal}\NormalTok{(}\FunctionTok{unname}\NormalTok{(test\_pca7\_eig}\SpecialCharTok{$}\NormalTok{vectors), }\FunctionTok{unname}\NormalTok{(test\_pca7}\SpecialCharTok{$}\NormalTok{loadings))}
\end{Highlighting}
\end{Shaded}

\begin{verbatim}
## [1] TRUE
\end{verbatim}

Confusingly, the function's output \texttt{eig} returns the square roots
of the eigenvalues:

\begin{Shaded}
\begin{Highlighting}[]
\FunctionTok{all.equal}\NormalTok{(}\FunctionTok{sqrt}\NormalTok{(test\_pca7\_eig}\SpecialCharTok{$}\NormalTok{values), test\_pca7}\SpecialCharTok{$}\NormalTok{eig)}
\end{Highlighting}
\end{Shaded}

\begin{verbatim}
## [1] TRUE
\end{verbatim}

In absolute values, the scores and loadings generated by \texttt{acp()}
are comparable to those obtained in the previous cases, as well as with
those in Tab.\ref{tab:tab5pdf}. Oddly, the output \texttt{sdev} of the
function \texttt{acp()} is computed directly as the St.Dev. of such
scores:

\begin{Shaded}
\begin{Highlighting}[]
\FunctionTok{all.equal}\NormalTok{(test\_pca7}\SpecialCharTok{$}\NormalTok{sdev, }\FunctionTok{apply}\NormalTok{(test\_pca7}\SpecialCharTok{$}\NormalTok{scores, }\DecValTok{2}\NormalTok{, sd)) }
\end{Highlighting}
\end{Shaded}

\begin{verbatim}
## [1] TRUE
\end{verbatim}

Yet the equivalence in Equation \ref{eq:eigvar} is not satisfied:

\begin{Shaded}
\begin{Highlighting}[]
\FunctionTok{all.equal}\NormalTok{(}\FunctionTok{unname}\NormalTok{(test\_pca7\_eig}\SpecialCharTok{$}\NormalTok{values), }\FunctionTok{apply}\NormalTok{(}\FunctionTok{unname}\NormalTok{(test\_pca7}\SpecialCharTok{$}\NormalTok{scores), }\DecValTok{2}\NormalTok{, var))}
\end{Highlighting}
\end{Shaded}

\begin{verbatim}
## [1] "Mean relative difference: 0.8"
\end{verbatim}

Unlike the implementations described so far, the function
\texttt{principal()} from the package
\emph{\href{https://CRAN.R-project.org/package=psych}{psych}}
(\protect\hyperlink{ref-psych}{Revelle, W. 2024}) runs an
eigendecomposition of the covariance matrix \(\mathbf{S}\), consistently
with our evaluation grid. Except that the meaning of ``loadings'' is
peculiarly defined as the eigenvectors rescaled by the square root of
the eigenvalues. This is reminiscent of---but not identical to---the
feature's coordinate in a biplot obtained in Equation \ref{eq:pcasd}.

If \texttt{principal()} is applied to the centred data matrix, and if
the implicit scaling just mentioned is corrected, then the scores are
equivalent, in absolute values, with those shown in
Tab.\ref{tab:tab5pdf}:

\begin{Shaded}
\begin{Highlighting}[]
\NormalTok{test\_pca8 }\OtherTok{\textless{}{-}}\NormalTok{ psych}\SpecialCharTok{::}\FunctionTok{principal}\NormalTok{(Y, }\AttributeTok{nfactors =} \DecValTok{0}\NormalTok{, }\AttributeTok{cor=}\StringTok{"cov"}\NormalTok{, }\AttributeTok{rotate =} \StringTok{"none"}\NormalTok{)}
\NormalTok{test\_pca8\_adjscor }\OtherTok{\textless{}{-}} \FunctionTok{sweep}\NormalTok{(}\FunctionTok{unname}\NormalTok{(test\_pca8}\SpecialCharTok{$}\NormalTok{scores), }\DecValTok{2}\NormalTok{, }\FunctionTok{sqrt}\NormalTok{(test\_pca8}\SpecialCharTok{$}\NormalTok{values), }\StringTok{"*"}\NormalTok{)}
\FunctionTok{all.equal}\NormalTok{(}\FunctionTok{abs}\NormalTok{(test\_pca8\_adjscor), }\FunctionTok{abs}\NormalTok{(}\FunctionTok{unname}\NormalTok{(Z)))}
\end{Highlighting}
\end{Shaded}

\begin{verbatim}
## [1] TRUE
\end{verbatim}

Similar considerations apply to the outputs \texttt{loadings} and
\texttt{values}. The latter returns the covariance matrix's eigenvalues,
and satisfies the equivalence in Equation \ref{eq:eigvar} provided that
the scores are once again corrected from their implicit scaling:

\begin{Shaded}
\begin{Highlighting}[]
\FunctionTok{all.equal}\NormalTok{(}\FunctionTok{unname}\NormalTok{(test\_pca8}\SpecialCharTok{$}\NormalTok{values), }\FunctionTok{apply}\NormalTok{(test\_pca8\_adjscor, }\DecValTok{2}\NormalTok{, var))}
\end{Highlighting}
\end{Shaded}

\begin{verbatim}
## [1] TRUE
\end{verbatim}

All three packages considered in this section offer visualisation
facilities for generating a combined plot of the PCA \emph{scores} and
\emph{loadings}. Examples include \texttt{scatter()} for
\texttt{dudi.pca()} objects; \texttt{plot()}, for \texttt{acp()}
objects; and \texttt{biplot()} for \texttt{principal()} objects. Yet
none of these functions seems to jointly compute the features' and
observations' coordinates according to Equation \ref{eq:svdbiplot},
hence one could question whether their visual output is a principal
component biplot.

What is more, objects generated by \texttt{dudi.pca()} can also be an
input to the function \texttt{ggbiplot()} discussed in the previous
section. A word of caution, however, is necessary as this biplot
implementation may not agree with Equation \ref{eq:svdbiplot} due to how
\texttt{get\_SVD()} chooses the equivalent of the left and right
eigenvector matrices needed to generate a biplot---in addition to the
caveats already mentioned.

\hypertarget{group-3-generalised-svd}{%
\subsection{Group 3: generalised SVD}\label{group-3-generalised-svd}}

The final group of implementations shares an intent to move beyond some
of the restrictions concerning what type of data can be handled by PCA,
allowing for the inclusion, for example, of ``mixed'' data. Another
commonality is that the underlying computational device is the so-called
``generalised'' SVD (\protect\hyperlink{ref-AbdiWilliams}{Abdi and
Williams 2010}, Appendix B). None of these characteristics seem to have
equivalent implementations in base-\texttt{R}.

The most prominent package in this group is probably
\emph{\href{https://CRAN.R-project.org/package=FactoMineR}{FactoMineR}}
(\protect\hyperlink{ref-factominerPaper}{Lê, Josse, and Husson 2008}),
whose functions \texttt{PCA()} is of particular interest here (e.g.,
\protect\hyperlink{ref-Pages2014MFAb}{Pagès 2014}, Ch. 1). This
implementation is underpinned by a generalised SVD function
\texttt{svd.triplet()}, which also operates under the hood of packages
such as
\emph{\href{https://CRAN.R-project.org/package=PCAmixdata}{PCAmixdata}}
(\protect\hyperlink{ref-chavent2022multivariate}{Chavent et al. 2022}).

As in previous cases, the PCA scores are equivalent, in absolute values,
to those in Tab.\ref{tab:tab5pdf}. Since the results of the generalised
SVD are available as part of the outputs of \texttt{PCA()}, users can
directly verify that the PCA scores are obtained consistently with
Equation \ref{eq:svdscores}:

\begin{Shaded}
\begin{Highlighting}[]
\NormalTok{ test\_pca9 }\OtherTok{\textless{}{-}}\NormalTok{ FactoMineR}\SpecialCharTok{::}\FunctionTok{PCA}\NormalTok{(X, }\AttributeTok{scale.unit =} \ConstantTok{FALSE}\NormalTok{)}
\end{Highlighting}
\end{Shaded}

\begin{Shaded}
\begin{Highlighting}[]
\NormalTok{D }\OtherTok{\textless{}{-}} \FunctionTok{diag}\NormalTok{(test\_pca9}\SpecialCharTok{$}\NormalTok{svd}\SpecialCharTok{$}\NormalTok{vs)}
\FunctionTok{all.equal}\NormalTok{(test\_pca9}\SpecialCharTok{$}\NormalTok{ind}\SpecialCharTok{$}\NormalTok{coord, test\_pca9}\SpecialCharTok{$}\NormalTok{svd}\SpecialCharTok{$}\NormalTok{U }\SpecialCharTok{\%*\%}\NormalTok{ D)}
\end{Highlighting}
\end{Shaded}

\begin{verbatim}
## [1] TRUE
\end{verbatim}

The singular values in this generalised SVD are related to the
eigenvalues of the covariance matrix whose elements are divided by
\(\sqrt{n}\) instead of \(\sqrt{n-1}\), which we encountered previously:

\begin{Shaded}
\begin{Highlighting}[]
\FunctionTok{all.equal}\NormalTok{(test\_pca9}\SpecialCharTok{$}\NormalTok{eig, test\_pca9}\SpecialCharTok{$}\NormalTok{svd}\SpecialCharTok{$}\NormalTok{vs}\SpecialCharTok{\^{}}\DecValTok{2}\NormalTok{)}
\end{Highlighting}
\end{Shaded}

\begin{verbatim}
## [1] TRUE
\end{verbatim}

\begin{Shaded}
\begin{Highlighting}[]
\FunctionTok{all.equal}\NormalTok{(test\_pca9}\SpecialCharTok{$}\NormalTok{eig, }\FunctionTok{unname}\NormalTok{(}\FunctionTok{eigen}\NormalTok{(}\DecValTok{1}\SpecialCharTok{/}\FunctionTok{nrow}\NormalTok{(Y)}\SpecialCharTok{*}\FunctionTok{t}\NormalTok{(Y) }\SpecialCharTok{\%*\%}\NormalTok{ Y)}\SpecialCharTok{$}\NormalTok{values))}
\end{Highlighting}
\end{Shaded}

\begin{verbatim}
## [1] TRUE
\end{verbatim}

However, this implementation too disagrees with the equivalence between
the scores' variance and covariance matrix's eigenvalues in Equation
\ref{eq:eigvar}:

\begin{Shaded}
\begin{Highlighting}[]
\FunctionTok{all.equal}\NormalTok{(}\FunctionTok{apply}\NormalTok{(test\_pca9}\SpecialCharTok{$}\NormalTok{ind}\SpecialCharTok{$}\NormalTok{coord,}\DecValTok{2}\NormalTok{,var) , test\_pca9}\SpecialCharTok{$}\NormalTok{eig)}
\end{Highlighting}
\end{Shaded}

\begin{verbatim}
## [1] "Mean relative difference: 0.1666667"
\end{verbatim}

Moving on to the biplot implementations in this group, it is worth
noting that \texttt{PCA()} generates coordinates for both features and
observations and it automatically produces a plot for each when called.
The observations features are simply the scores, whereas the features'
coordinates correspond to matrix \(\mathbf{B}\) in Equation
\ref{eq:pcasd}:

\begin{Shaded}
\begin{Highlighting}[]
\FunctionTok{all.equal}\NormalTok{(test\_pca9}\SpecialCharTok{$}\NormalTok{var}\SpecialCharTok{$}\NormalTok{coord, test\_pca9}\SpecialCharTok{$}\NormalTok{svd}\SpecialCharTok{$}\NormalTok{V }\SpecialCharTok{\%*\%}\NormalTok{ D)}
\end{Highlighting}
\end{Shaded}

\begin{verbatim}
## [1] TRUE
\end{verbatim}

Based on the above it seems reasonable to conclude that visualisation
devices like \texttt{FactoMineR::\ plot.PCA()} and
\texttt{PCAmixdata::plot.PCAmix()} do not generate, strictly speaking, a
principal component biplot. The package
\emph{\href{https://CRAN.R-project.org/package=factoextra}{factoextra}}
(\protect\hyperlink{ref-PCAalternatives}{Kassambara and Mundt 2020})
provides additional visualisation facilities building on the output
generated by \texttt{PCA()}, such as the function
\texttt{fviz\_pca\_biplot()}. Yet I could not ascertain an immediate
connection with Equation \ref{eq:svdbiplot}.

It is worth noting that \texttt{PCA()} and \texttt{fviz\_pca\_biplot()}
hint to the relationship between how features correlate and the angles
formed by the vectors that represent their coordinates on a biplot (see
e.g., \protect\hyperlink{ref-Pages2014MFAb}{Pagès 2014}, Ch. 1;
\protect\hyperlink{ref-STHDA_PCA}{Kassambara 2017}). Despite the
terminology used, these metrics bear little resemblance with the
properties concerning the the cosine of the angle between features'
vectors on a principal components biplot, or their lengths, from
Equations \ref{eq:length} and \ref{eq:coscorr}.

\hypertarget{concluding-remarks}{%
\section{Concluding remarks}\label{concluding-remarks}}

PCA and biplots are well-established techniques, and when it comes to
resources illustrating how they are done in practice, \texttt{R} users
are spoiled for choice. Whilst most resources hint to the underpinning
matrix decomposition approaches, it is not customary to dwell on how key
computational building blocks come about. Users are often left with a
rather procedural understanding of PCA and biplots tied to specific
implementations. Also, the overwhelming prevalence of SVD-based
approaches can obscure the rationale with which one arrives at such
concepts as \emph{scores} and \emph{loadings}.

In a context where implementing PCA and biplots has become somewhat a
conditioned reflex, one can hardly resist the temptation to give for
granted the internal workings. Yet in this note I have taken a
back-to-basics approach to comparing PCA and biplots implementations in
base \texttt{R} and a selection of contributed \texttt{R} packages. This
approach highlighted useful equivalences that should hold if the
computational rationale underpinning each technique is followed
correctly.

Findings suggest that discrepancies from an implementation-agnostic
understanding of PCA and biplots do arise, in both base \texttt{R} and
contributed \texttt{R} packages, from seemingly innocuous computational
choices made under the hood. Surprisingly, the identified verification
equivalences rarely follow without caveats from the output of specific
implementations alone. What is more, biplots are often just a misnomer
due to imprecisions in how the underlying computational aspects are
dealt with.

Base-\texttt{R} implementations such as \texttt{prcomp()} and
\texttt{biplot()} appear to be most convincingly aligned with what one
expects from the fundamental algebra of PCA and biplots, although some
caveats apply in the case of \texttt{biplot} that might call for minor
amendments. Despite being based on an SVD, functions like
\texttt{prcomp()} adhere more closely to our evaluation grid than
\texttt{princomp()} or \texttt{psych::principal()}. Functions with
richer capabilities and better visualisation devices, which are
understandably more popular amongst user, are harder to reconcile with
an implementation-agnostic understanding of PCA and biplots. This is the
case for packages like
\emph{\href{https://CRAN.R-project.org/package=FactoMineR}{FactoMineR}}
and \emph{\href{https://CRAN.R-project.org/package=ggbiplot}{ggbiplot}}.

This work has no pretence of comprehensiveness, and only some of the
scripts could be reviewed in an attempt to pinpoint possible
discrepancies from the proposed evaluation grid. Other resources provide
more hands-on comparative outlooks that might speak to practitioners
better than this note does; yet they rarely include biplots or offer an
implementation-agnostic perspective.

Despite its limitations, the hoped for outcome of this note is to raise
awareness that getting back-to-basic in PCA and biplot---two techniques
practically given for granted---helps to elevate aspects that are
usually disregarded for comparative purposes, and to address
discrepancies that users continue to find elusive despite the extensive
resources available for their implementation.

\hypertarget{references}{%
\section*{References}\label{references}}
\addcontentsline{toc}{section}{References}

\hypertarget{refs}{}
\begin{CSLReferences}{1}{0}
\leavevmode\vadjust pre{\hypertarget{ref-AbdiWilliams}{}}%
Abdi, H., and L. J. Williams. 2010. {``Principal Component Analysis.''}
\emph{WIREs Computational Statistics} 2 (4): 433--59.
https://doi.org/\url{https://doi.org/10.1002/wics.101}.

\leavevmode\vadjust pre{\hypertarget{ref-stackexchangePCA}{}}%
amoeba. 2015. {``Answer to: Making Sense of Principal Component
Analysis, Eigenvectors and Eigenvalues.''} Cross Validated. 2015.
\url{https://stats.stackexchange.com/q/140579}.

\leavevmode\vadjust pre{\hypertarget{ref-BinmoreDavies}{}}%
Binmore, K., and J. Davies. 2001. \emph{Calculus.} Cambridge: Cambridge
University Press.

\leavevmode\vadjust pre{\hypertarget{ref-PCAtools}{}}%
Blighe, K., and A. Lun. 2023. \emph{PCAtools: PCAtools: Everything
Principal Components Analysis}.
\url{https://doi.org/10.18129/B9.bioc.PCAtools}.

\leavevmode\vadjust pre{\hypertarget{ref-chemometric2014}{}}%
Bro, R., and A. K. Smilde. 2014. {``Principal Component Analysis.''}
\emph{Anal. Methods} 6: 2812--31.
\url{https://doi.org/10.1039/C3AY41907J}.

\leavevmode\vadjust pre{\hypertarget{ref-chavent2022multivariate}{}}%
Chavent, M., V. Kuentz-Simonet, A. Labenne, and J. Saracco. 2022.
{``Multivariate Analysis of Mixed Data: The r Package PCAmixdata.''}
\url{https://arxiv.org/abs/1411.4911}.

\leavevmode\vadjust pre{\hypertarget{ref-DuToit1986GEDA}{}}%
du Toit, S. H. C., A. G. W. Steyn, and R. H. Stumpf. 1986.
\emph{Graphical Exploratory Data Analysis}. Springer.

\leavevmode\vadjust pre{\hypertarget{ref-EverittBrian2011Aita}{}}%
Everitt, B., and T. Hothorn. 2011. \emph{An Introduction to Applied
Multivariate Analysis with r}. Springer.

\leavevmode\vadjust pre{\hypertarget{ref-Gower2011}{}}%
Gower, J. C., S. Lubbe, and N. LeRoux. 2011. \emph{Understanding
Biplots}. Chichester: Wiley.

\leavevmode\vadjust pre{\hypertarget{ref-LearnPCA}{}}%
Hanson, Bryan A., and David T. Harvey. 2022. \emph{LearnPCA: Functions,
Data Sets and Vignettes to Aid in Learning Principal Components Analysis
(PCA)}. \url{https://CRAN.R-project.org/package=LearnPCA}.

\leavevmode\vadjust pre{\hypertarget{ref-JolliffeI.T.2004Pca}{}}%
Jolliffe, I. T. 2004. \emph{Principal Component Analysis}. 2nd ed.
Springer.

\leavevmode\vadjust pre{\hypertarget{ref-STHDA_PCA}{}}%
Kassambara, A. 2017. {``PCA - Principal Component Analysis
Essentials.''} STHDA. 2017.
\url{http://www.sthda.com/english/articles/31-principal-component-methods-in-r-practical-guide/112-pca-principal-component-analysis-essentials/\#r-packages}.

\leavevmode\vadjust pre{\hypertarget{ref-PCAalternatives}{}}%
Kassambara, A., and F. Mundt. 2020. \emph{Factoextra : Extract and
Visualize the Results of Multivariate Data Analyses}.
\url{https://cran.r-project.org/web/packages/factoextra/readme/README.html}.

\leavevmode\vadjust pre{\hypertarget{ref-KnoxStevenW.2018Ml:t}{}}%
Knox, S. W. 2018. \emph{Machine Learning : Topics and Techniques}. 1st
ed. Wiley.

\leavevmode\vadjust pre{\hypertarget{ref-Kumar2016Mtmw}{}}%
Kumar, A., and A. Paul. 2016. \emph{Mastering Text Mining with r}. Pakt
Publishing.

\leavevmode\vadjust pre{\hypertarget{ref-factominerPaper}{}}%
Lê, S., J. Josse, and F. Husson. 2008. {``{FactoMineR}: A Package for
Multivariate Analysis.''} \emph{Journal of Statistical Software} 25 (1):
1--18. \url{https://doi.org/10.18637/jss.v025.i01}.

\leavevmode\vadjust pre{\hypertarget{ref-amap}{}}%
Lucas, A. 2022. \emph{Amap: Another Multidimensional Analysis Package}.
\url{https://CRAN.R-project.org/package=amap}.

\leavevmode\vadjust pre{\hypertarget{ref-intermedAlgebra}{}}%
Marecek, L. 2017. \emph{Intermediate Algebra.} OpenStax.
\url{https://openstax.org/books/intermediate-algebra/pages/preface}.

\leavevmode\vadjust pre{\hypertarget{ref-MayorEric2015Lpaw}{}}%
Mayor, E. 2015. \emph{Learning Predictive Analytics with r}. Pakt
Publishing.

\leavevmode\vadjust pre{\hypertarget{ref-Pages2014MFAb}{}}%
Pagès, J. 2014. \emph{Multiple Factor Analysis by Example Using r}. 1st
ed. Chapman; Hall/CRC.

\leavevmode\vadjust pre{\hypertarget{ref-Peng_PCA}{}}%
Peng, R. D. 2020. {``Exploratory Data Analysis with r - SVD and PCA.''}
2020.
\url{https://bookdown.org/rdpeng/exdata/dimension-reduction.html\#svd-and-pca}.

\leavevmode\vadjust pre{\hypertarget{ref-PooleDavid2015}{}}%
Poole, D. 2014. \emph{Linear Algebra : A Modern Introduction}. 4th ed.
Brooks Cole.

\leavevmode\vadjust pre{\hypertarget{ref-psych}{}}%
Revelle, W. 2024. \emph{Psych: Procedures for Psychological,
Psychometric, and Personality Research}. Evanston, Illinois:
Northwestern University. \url{https://CRAN.R-project.org/package=psych}.

\leavevmode\vadjust pre{\hypertarget{ref-pcaMethods_paper}{}}%
Stacklies, W., H. Redestig, M. Scholz, D. Walther, and J. Selbig. 2007.
{``pcaMethods -- a Bioconductor Package Providing PCA Methods for
Incomplete Data.''} \emph{Bioinformatics} 23: 1164--67.
\url{https://doi.org/10.1093/bioinformatics/btm069}.

\leavevmode\vadjust pre{\hypertarget{ref-StatQuestPCA}{}}%
Starmer, J. 2018. {``StatQuest: Principal Component Analysis (PCA),
Step-by-Step.''} {[}media{]} Youtube. 2018.
\url{https://youtu.be/FgakZw6K1QQ?si=ROWqV5NwFNNKCXVv}.

\leavevmode\vadjust pre{\hypertarget{ref-ade4book}{}}%
Thioulouse, J., S. Dray, Dufour A.--B, A. Siberchicot, T. Jombart, and
S. Pavoine. 2018. \emph{Multivariate Analysis of Ecological Data with
{ade4}}. Springer.

\leavevmode\vadjust pre{\hypertarget{ref-Venables.2002Masw}{}}%
Venables, W. N., and B. D. Ripley. 2002. \emph{Modern Applied Statistics
with s}. 4th ed. Springer.

\leavevmode\vadjust pre{\hypertarget{ref-ggbiplot_vignette}{}}%
Vu, V. Q., and M. Friendly. 2024. \emph{Ggbiplot: A Grammar of Graphics
Implementation of Biplots}.
\url{https://CRAN.R-project.org/package=ggbiplot}.

\end{CSLReferences}


\end{document}